\documentclass[11pt]{article}
\usepackage{graphicx}
\usepackage{amsmath}
\usepackage{amssymb}
\usepackage{caption2}
\begin{document}
\bibliographystyle{prsty}
\begin{center}
{\large {\bf \sc{  Analysis of $Y(4660)$ and related bound states  with   QCD sum rules}}} \\[2mm]
Zhi-Gang Wang \footnote{wangzgyiti@yahoo.com.cn.  }, Xiao-Hong Zhang     \\
 Department of Physics, North China Electric Power University,
Baoding 071003, P. R. China
\end{center}

\begin{abstract}
In this article, we take the vector charmonium-like state  $Y(4660)$
as a $\psi'f_0(980)$ bound state (irrespective of the
hadro-charmonium and the molecular state) tentatively, study its
mass  using the QCD sum rules, the numerical value
$M_Y=4.71\pm0.26\,\rm{GeV}$ is consistent with the experimental
data. Considering the $SU(3)$ symmetry of the light flavor quarks
and the heavy quark symmetry, we also study the bound states
$\psi'\sigma(400-1200)$, $\Upsilon'''f_0(980)$ and
$\Upsilon'''\sigma(400-1200)$ with the QCD sum rules, and make
reasonable predictions for their masses.
\end{abstract}



\section{Introduction}
  Two resonant structures are observed in the $\pi^+\pi^-
\psi'$ invariant mass distribution in the  cross section for the
process $e^+ e^- \to \pi^+\pi^- \psi'$ between threshold and
$\sqrt{s}=5.5~\rm{GeV}$
 using $673~\rm{fb}^{-1}$ of data on and off the
$\Upsilon(4S)$ ($\Upsilon'''$) resonance collected with the Belle
detector at KEK-B, one at  $4361\pm 9\pm 9~\rm{MeV}$ with a width of
$74\pm 15\pm 10~\rm{MeV}$, and another at $4664\pm 11\pm 5~\rm{MeV}$
with a width of $48\pm 15\pm 3~\rm{MeV}$ (they are denoted as $Y
(4360)$ and $Y(4660)$ respectively), where the mass spectrum is
parameterized by two Breit-Wigner functions \cite{Yexp}. The
structure $Y(4660)$ is neither observed in the initial state
radiation (ISR) process $e^+e^-\to\gamma_{ISR}\pi^+\pi^-J/\psi$
\cite{ISRJpsi}, nor in the exclusive cross processes  $e^+e^-\to
D{\bar D},D{\bar D}^*, D^*{\bar D}^*, D{\bar D}\pi$, $ J/\psi
D^{(*)} {\bar D}^{(*)}$
\cite{ExpDDbar1,ExpDDbar2,ExpDDbar3,ExpDDbar4,Abe:2007sy}.

There have been several  canonical charmonium interpretations for
the $Y(4660)$, such as the $5^3S_1$ state \cite{Ding0708}, the
$6^3S_1$ state \cite{Chao0903}, the $5^3S_1-4^3D_1$ mixing state
\cite{SDmix}. In Ref.\cite{Qiao0709}, Qiao suggests that the
$Y(4660)$ is  a baryonium state, the radial excited state of the
$\frac{1}{\sqrt{2}}(|\Lambda_c\bar{\Lambda}_c\rangle+|\Sigma^0_c\bar{\Sigma}^0_c\rangle)$.
In Ref.\cite{Nielsen0804}, Albuquerque et al take the $Y(4660)$ as a
vector $cs\bar{c}\bar{s}$ tetraquark state, and study its mass with
the QCD sum rules.
 A critical information for understanding the structure of those  charmonium-like states is
wether or not the $\pi\pi$ comes from a resonance. There is some
indication that only the $Y(4660)$ has a well defined intermediate
state which is consistent with the scalar meson $f_0(980)$ in the
$\pi\pi$ invariant mass spectra \cite{babarconf0801}.  In
Ref.\cite{Guo0803}, Guo et al take the $Y(4660)$ as a
$\psi'f_0(980)$ bound state (molecular state) considering   the
nominal threshold of the $\psi'-f_0(980)$ system is about
$4666\pm10$~MeV \cite{PDG}, the $Y(4660)$  decays  dominantly via
the decay of the scalar meson $f_0(980)$,
$Y(4660)\to\psi'f_0(980)\to \psi'\pi\pi$, $ \psi'K{\bar K}$, the
difficulties in the canonical charmonium interpretation can be
overcome.
 In Refs.\cite{Voloshin0803,VoloshinReV}, Voloshin et al argue that the
charmonium-like states $Y(4660)$, $Z(4430)$, $Y(4260)$, $\cdots$ may
be hadro-charmonia. The relatively compact charmonium states
($J/\psi$, $\psi'$ and $\chi_{cJ}$) can be bound inside light
hadronic matter, in particular inside higher resonances made from
light quarks and (or) gluons. The charmonium state in such binding
 retains its properties essentially,  the bound system
(hadro-charmonium, a special molecular state) decays into light
mesons and the particular charmonium.

In this article, we study the mass of the $Y(4660)$ as a
$\psi'f_0(980)$ bound state (irrespective of the hadro-charmonium
and the molecular state) using the QCD sum rules
\cite{SVZ79,Reinders85}. As a byproduct, we take into account the
$SU(3)$ symmetry of the light flavor quarks and the heavy quark
symmetry, study the related hidden charm and hidden bottom states.
In the QCD sum rules, the operator product expansion is used to
expand the time-ordered currents into a series of quark and gluon
condensates which parameterize the long distance properties of the
QCD vacuum. Based on the quark-hadron duality, we can obtain copious
information about the hadronic parameters at the phenomenological
side \cite{SVZ79,Reinders85}.

The article is arranged as follows:  we derive the QCD sum rules for
  the vector charmonium-like state $Y(4660)$ and the related bound  states
  in section 2; in section 3, numerical
results and discussions; section 4 is reserved for conclusion.

\section{QCD sum rules for  the $Y(4660)$ and related bound  states }
In the following, we write down  the two-point correlation functions
$\Pi_{\mu\nu}(p)$  in the QCD sum rules,
\begin{eqnarray}
\Pi_{\mu\nu}(p)&=&i\int d^4x e^{ip \cdot x} \langle
0|T\left[J/\eta_\mu(x)J/\eta_\nu^{\dagger}(0)\right]|0\rangle \, , \\
J_\mu(x)&=&\bar{Q}(x)\gamma_\mu Q(x) \bar{s}(x)s(x) \, , \nonumber\\
\eta_\mu(x)&=&\frac{1}{\sqrt{2}}\bar{Q}(x)\gamma_\mu Q(x)
\left[\bar{u}(x)u(x)+\bar{d}(x)d(x) \right]\, ,
\end{eqnarray}
where the $Q$ denotes the heavy quarks $c$ and $b$. We use the
currents $J_\mu(x)$ and $\eta_\mu(x)$ ($Q=c$) to interpolate the
bound states $\psi'f_0(980)$ and $\psi'\sigma(400-1200)$,
respectively. The $Y(4660)$ can be tentatively identified as the
$\psi'f_0(980)$ bound state, while there lack experimental
candidates to identify the $\psi'\sigma(400-1200)$ bound state.
Considering the heavy quark symmetry, there maybe exist some hidden
bottom  bound states, for example, $\Upsilon f_0(980)$, $\Upsilon'
f_0(980)$, $\Upsilon'' f_0(980)$, $\Upsilon''' f_0(980)$,
$\Upsilon\sigma(400-1200)$, $\Upsilon'\sigma(400-1200)$,
$\Upsilon''\sigma(400-1200)$, $\Upsilon'''\sigma(400-1200)$,
$\cdots$, we study those possibilities  with the  currents
$J_\mu(x)$ and $\eta_\mu(x)$ ($Q=b$), and make predictions for their
masses which are fundamental parameters in describing a hadron.

The hidden charm current $\bar{c}(x)\gamma_\mu c(x)$ can interpolate
the charmonia $J/\psi$, $\psi'$, $\psi(3770)$, $\psi(4040)$,
$\psi(4160)$, $\psi(4415)$, $\cdots$; while the hidden bottom
current $\bar{b}(x)\gamma_\mu b(x)$ can interpolate the bottomonia
$\Upsilon$, $\Upsilon'$, $\Upsilon''$, $\Upsilon'''$,
$\Upsilon''''$, $\cdots$ \cite{PDG}. We assume that the scalar
mesons $f_0(980)$ and $\sigma(400-1200)$ are the conventional
$q\bar{q}$ states, to be more precise, they have large $q\bar{q}$
components. The currents $J_\mu(x)$ and $\eta_\mu(x)$ ($Q=c$) have
non-vanishing couplings with the bound states $J/\psi f_0(980)$,
$\psi'f_0(980)$, $\psi''f_0(980)$, $\cdots$ and
$J/\psi\sigma(400-1200)$, $\psi'\sigma(400-1200)$,
$\psi''\sigma(400-1200)$, $\cdots$, respectively. The colored
objects (diquarks) in a  confining potential can result in a copious
spectrum, there maybe  exist  a series of orbital angular momentum
excitations; while the colorless objects (mesons) bound by a short
range potential (through meson-exchange) should have a very limited
spectrum, it is relatively easy to identify the molecule type bound
states. We determine the masses of the ground states by imposing the
two criteria  of the QCD sum rules, then compare them with the
nominal thresholds of the corresponding systems $J/\psi-f_0(980)$,
$\psi'-f_0(980)$, $\cdots$. In Ref.\cite{Voloshin0803}, Voloshin et
al argue that a formation of hadro-charmonium is favored for higher
charmonium resonances $\psi'$ and $\chi_{cJ}$ as compared to the
lowest states $J/\psi$ and $\eta_c$.

We can insert  a complete set of intermediate hadronic states with
the same quantum numbers as the current operators $J_\mu(x)$ and
$\eta_\mu(x)$ into the correlation functions  $\Pi_{\mu\nu}(p)$  to
obtain the hadronic representation \cite{SVZ79,Reinders85}. After
isolating the ground state contributions  from the pole terms of the
$Y$ and $Z$, we get the following result,
\begin{eqnarray}
\Pi_{\mu\nu}(p)&=&\frac{\lambda_{Y}^2}{M_{Y}^2-p^2}\left[
-g_{\mu\nu}+\frac{p_\mu p_\nu}{p^2}\right]
+\frac{\lambda_Z^2}{M_{Z}^2-p^2}p_\mu p_\nu +\cdots \, \, ,
\end{eqnarray}
where the pole residues (or coupling) $\lambda_Y$ and $\lambda_Z$
are defined by
\begin{eqnarray}
\lambda_{Y}\epsilon_\mu  &=& \langle 0|J/\eta_\mu(0)|Y(p)\rangle \, ,\nonumber \\
\lambda_{Z}p_\mu  &=& \langle 0|J/\eta_\mu(0)|Z(p)\rangle \, ,
\end{eqnarray}
and the $\epsilon_\mu$ is the  polarization  vector. In Eq.(3), we
show the contribution from the scalar bound state $Z$ explicitly,
because the vector currents $J_\mu(x)$ and $\eta_\mu(x)$ are by no
means conserved.

 After performing the standard procedure of the QCD sum rules, we obtain   two  sum rules for the
 $c\bar{c}s\bar{s}$ and $b\bar{b}s\bar{s}$ channels respectively (In the isospin
limit, the interpolating currents result in two distinct expressions
for the correlation functions $\Pi_{\mu\nu}(p)$, which are
characterized by the number of the $s$ quark they contain,
thereafter will use the quark constituents  to denote the
corresponding quantities.):
\begin{eqnarray}
\lambda_{Y}^2 e^{-\frac{M_Y^2}{M^2}}= \int_{\Delta}^{s_0} ds
\rho(s)e^{-\frac{s}{M^2}} \, ,
\end{eqnarray}
\begin{eqnarray}
\rho(s)&=&\rho_{0}(s)+\rho_{\langle
\bar{s}s\rangle}(s)+\left[\rho^A_{\langle
GG\rangle}(s)+\rho^B_{\langle GG\rangle}(s)\right]\langle
\frac{\alpha_s GG}{\pi}\rangle+\rho_{\langle \bar{s}s\rangle^2}(s)\,
.
\end{eqnarray}
The explicit expressions of the spectral densities $\rho_{0}(s)$,
$\rho_{\langle \bar{s}s\rangle}(s)$, $\rho^A_{\langle
GG\rangle}(s)$, $\rho^B_{\langle GG\rangle}(s)$ and $\rho_{\langle
\bar{s}s\rangle^2}(s)$ are presented in  the appendix.
      The $s_0$ is the continuum threshold parameter and the $M^2$ is the Borel parameter;
      $\alpha_{f}=\frac{1+\sqrt{1-4m_Q^2/s}}{2}$,
$\alpha_{i}=\frac{1-\sqrt{1-4m_Q^2/s}}{2}$, $\beta_{i}=\frac{\alpha
m_Q^2}{\alpha s -m_Q^2}$,
$\widetilde{m}_Q^2=\frac{(\alpha+\beta)m_Q^2}{\alpha\beta}$,
$\widetilde{\widetilde{m}}_Q^2=\frac{m_Q^2}{\alpha(1-\alpha)}$, and
$\Delta=4(m_Q+m_s)^2$. We can obtain  two sum rules for the
$c\bar{c}q\bar{q}$ and $b\bar{b}q\bar{q}$ channels with a simple
replacement $m_s\rightarrow 0$,  $\langle
\bar{s}s\rangle\rightarrow\langle \bar{q}q\rangle$ and $\langle
\bar{s}g_s \sigma Gs\rangle\rightarrow\langle \bar{q}g_s \sigma
Gq\rangle$.

 We carry out the operator
product expansion (OPE) to the vacuum condensates adding up to
dimension-10. In calculation, we
 take  assumption of vacuum saturation for  high
dimension vacuum condensates, they  are always
 factorized to lower condensates with vacuum saturation in the QCD sum rules,
  factorization works well in  the large $N_c$ limit. Moreover, we neglect the terms
proportional to the $m_u$ and $m_d$, their contributions are of
minor importance.

 Differentiate  the Eq.(5) with respect to  $\frac{1}{M^2}$, then eliminate the
 pole residue $\lambda_{Y}$, we can obtain the  sum rule for
 the mass  of the bound state $Y$,
 \begin{eqnarray}
 M_Y^2= \frac{\int_{\Delta}^{s_0} ds
\frac{d}{d(-1/M^2)}\rho(s)e^{-\frac{s}{M^2}} }{\int_{\Delta}^{s_0}
ds \rho(s)e^{-\frac{s}{M^2}}}\, .
\end{eqnarray}

\section{Numerical results and discussions}
The input parameters are taken to be the standard values $\langle
\bar{q}q \rangle=-(0.24\pm 0.01 \,\rm{GeV})^3$, $\langle \bar{s}s
\rangle=(0.8\pm 0.2 )\langle \bar{q}q \rangle$, $\langle
\bar{q}g_s\sigma Gq \rangle=m_0^2\langle \bar{q}q \rangle$, $\langle
\bar{s}g_s\sigma Gs \rangle=m_0^2\langle \bar{s}s \rangle$,
$m_0^2=(0.8 \pm 0.2)\,\rm{GeV}^2$, $\langle \frac{\alpha_s
GG}{\pi}\rangle=(0.33\,\rm{GeV})^4 $, $m_s=(0.14\pm0.01)\,\rm{GeV}$,
$m_c=(1.35\pm0.10)\,\rm{GeV}$ and $m_b=(4.8\pm0.1)\,\rm{GeV}$ at the
energy scale   $\mu=1\, \rm{GeV}$ \cite{SVZ79,Reinders85,Ioffe2005}.

In the conventional QCD sum rules \cite{SVZ79,Reinders85}, there are
two criteria (pole dominance and convergence of the operator product
expansion) for choosing  the Borel parameter $M^2$ and threshold
parameter $s_0$.  We impose the two criteria on the charmonium-like
states $Y$ to choose the Borel parameter $M^2$ and threshold
parameter $s_0$. The light tetraquark states cannot satisfy the two
criteria, although it is not an indication non-existence of the
light tetraquark states (For detailed discussions about this
subject, one can consult Refs.\cite{Wang08072,Wang0708}).

We take the vector charmonium-like state $Y(4660)$ as the
$\psi'f_0(980)$ bound state tentatively,  and take the threshold
parameter as $s^0_{s\bar{s}}=(4.66+0.5)^2\, \rm{GeV}^2\approx 27 \,
\rm{GeV}^2$    to take into account  possible contribution from the
ground state,  where we choose the energy gap between the ground
state and the first radial excited state to be $0.5\,\rm{GeV}$.
Taking into account the $SU(3)$ symmetry of the light flavor quarks,
we expect the threshold parameter $s^0_{q\bar{q}}$ (for the bound
state $\psi'\sigma(400-1200)$) is slightly smaller than  the
$s^0_{s\bar{s}}$. Furthermore, we take into account the mass
difference between the $c$ and $b$ quarks, the threshold parameters
in the   hidden bottom channels are tentatively taken as
$s^0_{q\bar{q}}=144\, \rm{GeV}^2$ and $s^0_{s\bar{s}}=145\,
\rm{GeV}^2$.

In this article,  we take it for granted that the energy gap between
the ground state and the first radial excited state is about
$0.5\,\rm{GeV}$, and use this  value as a guide to determine the
threshold parameter $s_0$ with the QCD sum rules.

The contributions from the high dimension vacuum condensates  in the
operator product expansion are shown in Figs.1-2, where (and
thereafter) we  use the $\langle\bar{q}q\rangle$ to denote the quark
condensates $\langle\bar{q}q\rangle$, $\langle\bar{s}s\rangle$ and
the $\langle\bar{q}g_s \sigma Gq\rangle$ to denote the mixed
condensates $\langle\bar{q}g_s \sigma Gq\rangle$, $\langle\bar{s}g_s
\sigma Gs\rangle$. From the figures, we can see that the
contributions from the high dimension condensates change  quickly
with variation of the Borel parameter at the values $M^2\leq
2.8\,\rm{GeV}^2$ and $M^2\leq 7.5\,\rm{GeV}^2$ for the hidden charm
 and hidden  bottom channels respectively, such an unstable behavior
cannot lead to stable sum rules, our numerical results confirm this
conjecture, see Fig.4.

At the values $M^2\geq 2.8\,\rm{GeV}^2$ and $s_0\geq
26\,\rm{GeV}^2$, the contributions from the  $\langle
\bar{q}q\rangle^2+\langle \bar{q}q\rangle \langle \bar{q}g_s \sigma
Gq\rangle $ term are less than (or equal) $18.5\%$ for the
$c\bar{c}s\bar{s}$ channel, the corresponding contributions are less
than (or equal) $ 36.5\%$ for the $c\bar{c}q\bar{q}$  channel; the
contributions from the vacuum condensate of the highest dimension
$\langle\bar{q}g_s \sigma Gq\rangle^2$ are less than $5\%$ for all
the hidden charm channels, we expect the operator product expansion
is convergent in the hidden charm channels. At the values $M^2\geq
7.6\,\rm{GeV}^2$ (In Figs.2-4, the vertical line corresponds to the
value  $M^2=7.6\,\rm{GeV}^2$ in the hidden bottom channels.) and
$s_0\geq 148\,\rm{GeV}^2$, the contributions from the $\langle
\bar{q}q\rangle^2+\langle \bar{q}q\rangle \langle \bar{q}g_s \sigma
Gq\rangle $ term are less than  $7\%$ for the $b\bar{b}s\bar{s}$
channel, the corresponding contributions are less than (or equal) $
18\%$ for the $b\bar{b}q\bar{q}$  channel; the contributions from
the vacuum condensate of the highest dimension $\langle\bar{q}g_s
\sigma Gq\rangle^2$ are less than (or equal) $7\%$ for all the
hidden bottom  channels, we expect the operator product expansion is
convergent in the hidden bottom channels.

The contributions from the  gluon condensate $\langle \frac{\alpha_s
GG}{\pi}\rangle$ are rather large, while the contributions from the
high dimension condensates $\langle \frac{\alpha_s GG}{\pi} \rangle
\left[\langle \bar{q} q \rangle +\langle \bar{q}g_s\sigma G q
\rangle+ \langle \bar{q} q \rangle^2\right]$ are small enough, the
total contributions involving the gluon condensate are less than (or
equal) $30\%$ ($22\%$) for the $c\bar{c}s\bar{s}$
($c\bar{c}q\bar{q}$) channel at the values $M^2\geq 2.8\,\rm{GeV}^2$
and $s_0\geq 26\,\rm{GeV}^2$; while the  contributions  are less
than $21\%$ ($17\%$) for the $b\bar{b}s\bar{s}$ ($b\bar{b}q\bar{q}$)
channel at the values $M^2\geq 7.6\,\rm{GeV}^2$ and $s_0\geq
148\,\rm{GeV}^2$. In the QCD sum rules for the tetraquark states
(irrespective of the molecule type and the diquark-antidiquark
type), the contributions from the gluon condensate are suppressed by
large denominators and would not play any significant roles for the
light tetraquark states \cite{Wang1,Wang2}, the heavy tetraquark
state \cite{Wang08072} and the  heavy molecular state
\cite{Wang0904}; the present sum rules seem rather exotic. If we
take a simple replacement $\bar{s}(x)s(x)\rightarrow \langle
\bar{s}s\rangle$ and $\left[\bar{u}(x)u(x)+\bar{d}(x)d(x)
\right]\rightarrow 2\langle\bar{q}q\rangle$ in the interpolating
currents $J_\mu(x)$ and $\eta_\mu(x)$, we can obtain the standard
vector heavy quark current $Q(x)\gamma_\mu Q(x)$, where the gluon
condensate plays an important rule in the QCD sum rules
\cite{SVZ79}.

In calculation, we observe that the dominant contributions come from
the perturbative term and the $\langle \bar{q}q\rangle+ \langle
\bar{q}g_s \sigma Gq\rangle $ term at the values $M^2\geq
2.8\,\rm{GeV}^2$ and $s_0\geq 26\,\rm{GeV}^2$ for the hidden charm
channels and at the values $M^2\geq 7.6\,\rm{GeV}^2$ and $s_0\geq
148\,\rm{GeV}^2$ for the hidden bottom channels, the operator
product expansion is convergent.

 In this article, we take the uniform Borel parameter
$M^2_{min}$, i.e. $M^2_{min}\geq 2.8 \, \rm{GeV}^2$ and
$M^2_{min}\geq 7.6 \, \rm{GeV}^2$ for the hidden charm  and hidden
bottom channels, respectively.

\begin{figure}
 \centering
 \includegraphics[totalheight=4.5cm,width=6cm]{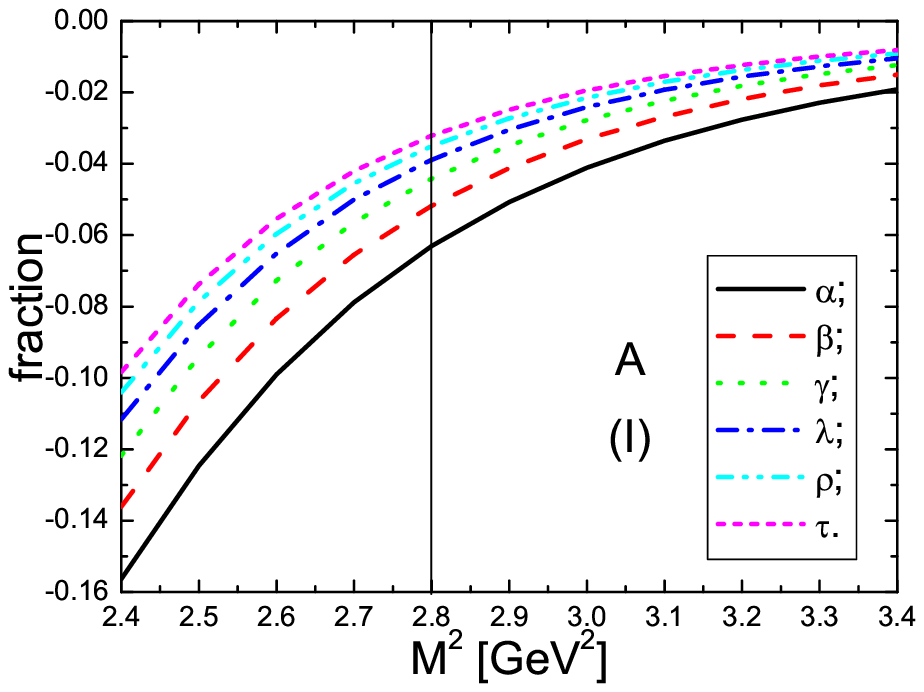}
 \includegraphics[totalheight=4.5cm,width=6cm]{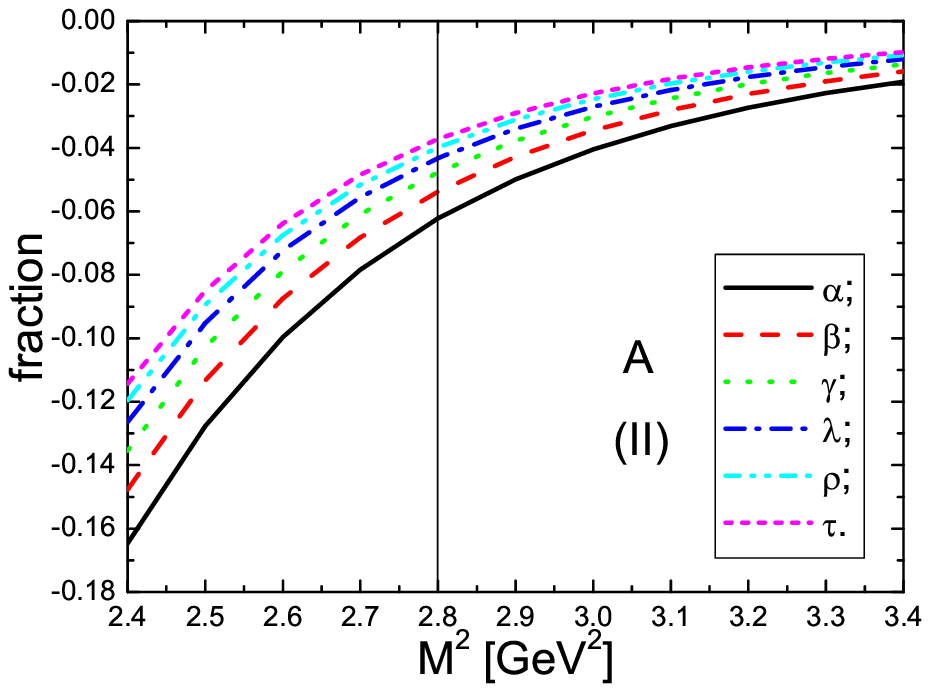}
 \includegraphics[totalheight=4.5cm,width=6cm]{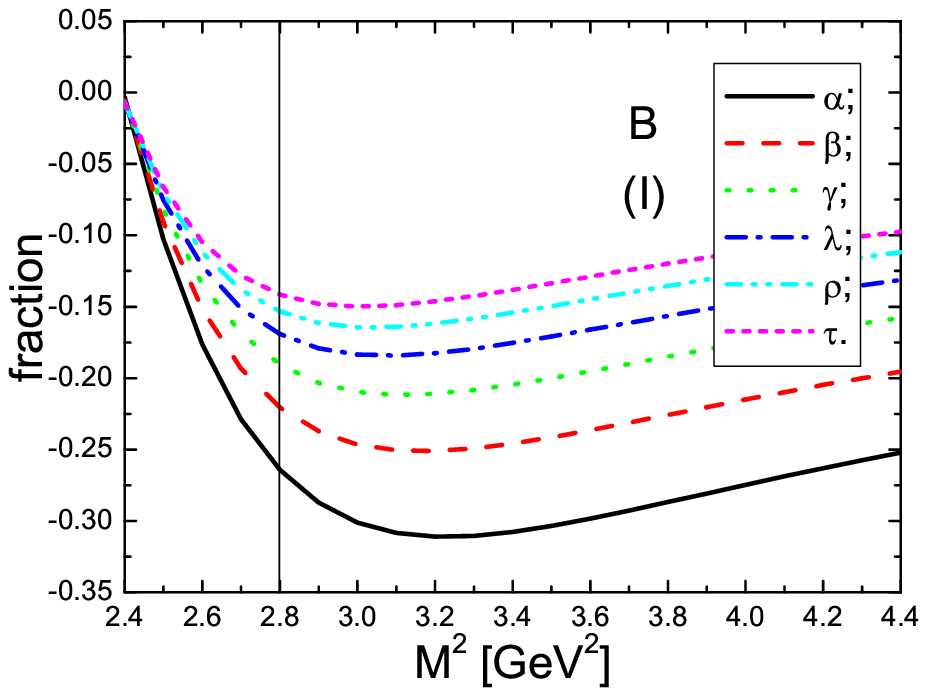}
 \includegraphics[totalheight=4.5cm,width=6cm]{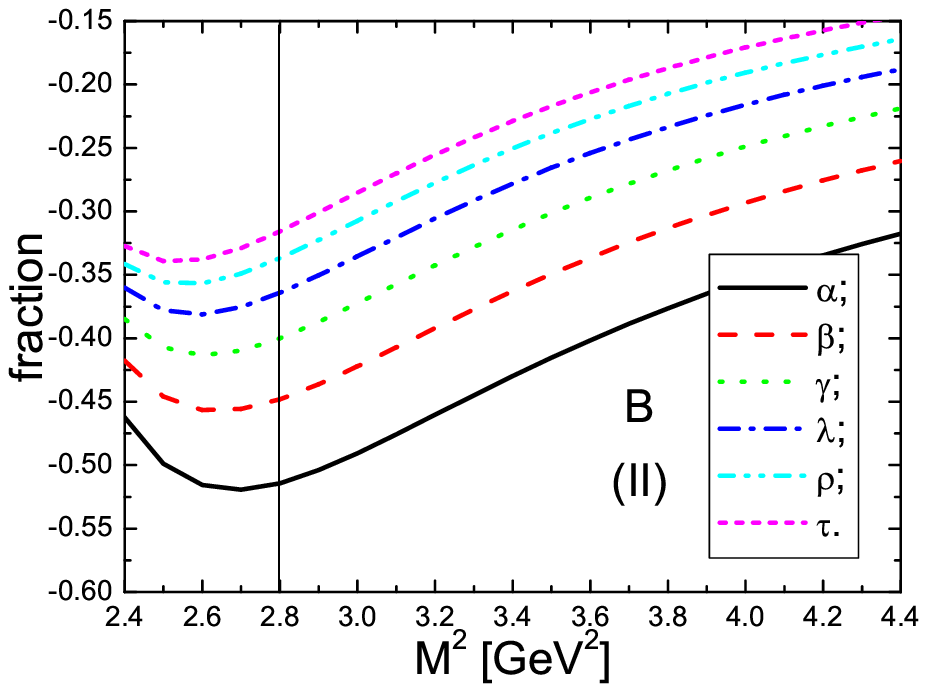}
 \includegraphics[totalheight=4.5cm,width=6cm]{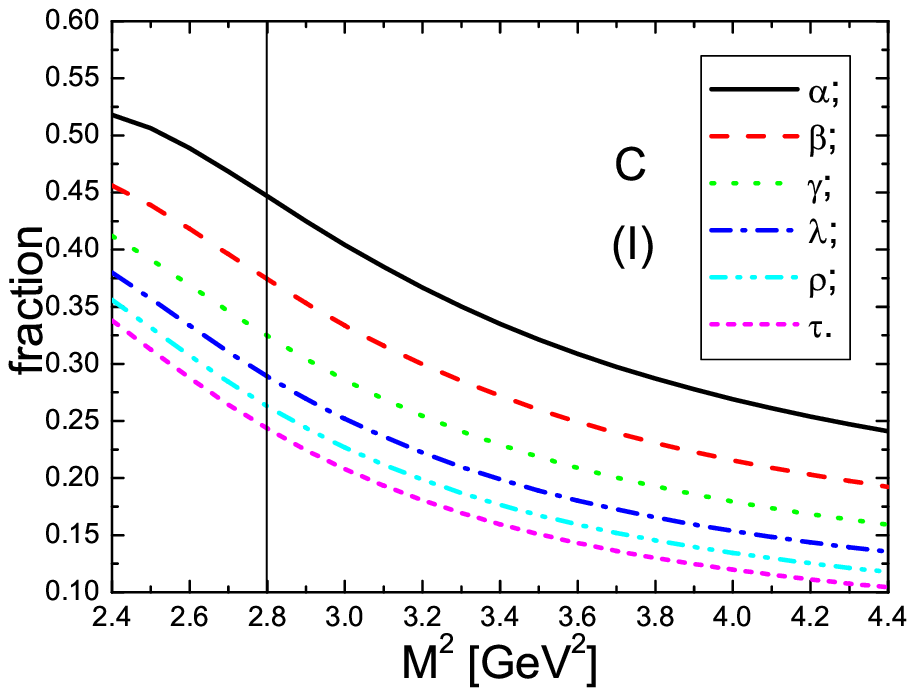}
 \includegraphics[totalheight=4.5cm,width=6cm]{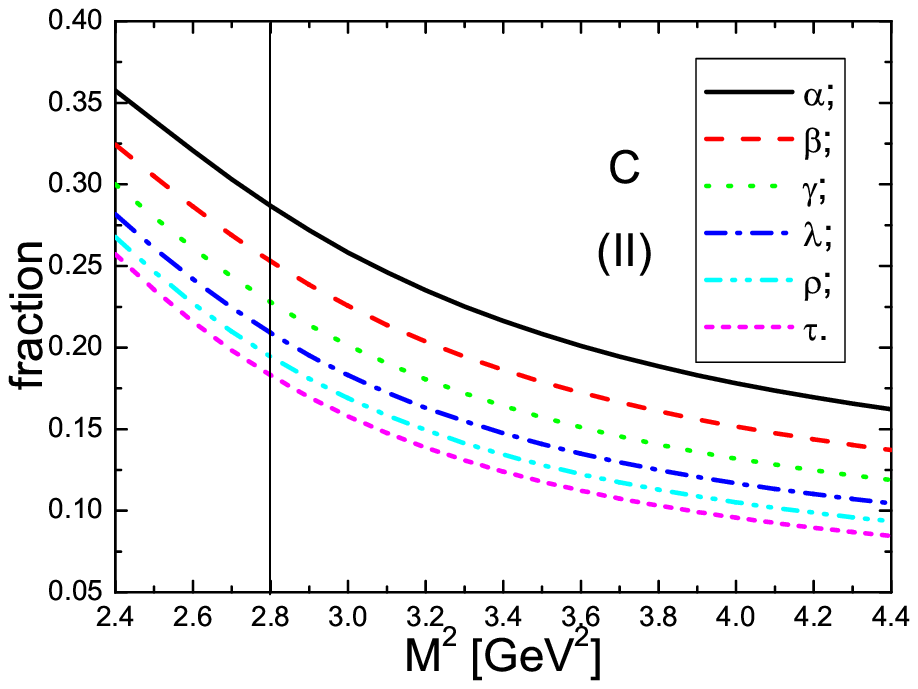}
  \includegraphics[totalheight=4.5cm,width=6cm]{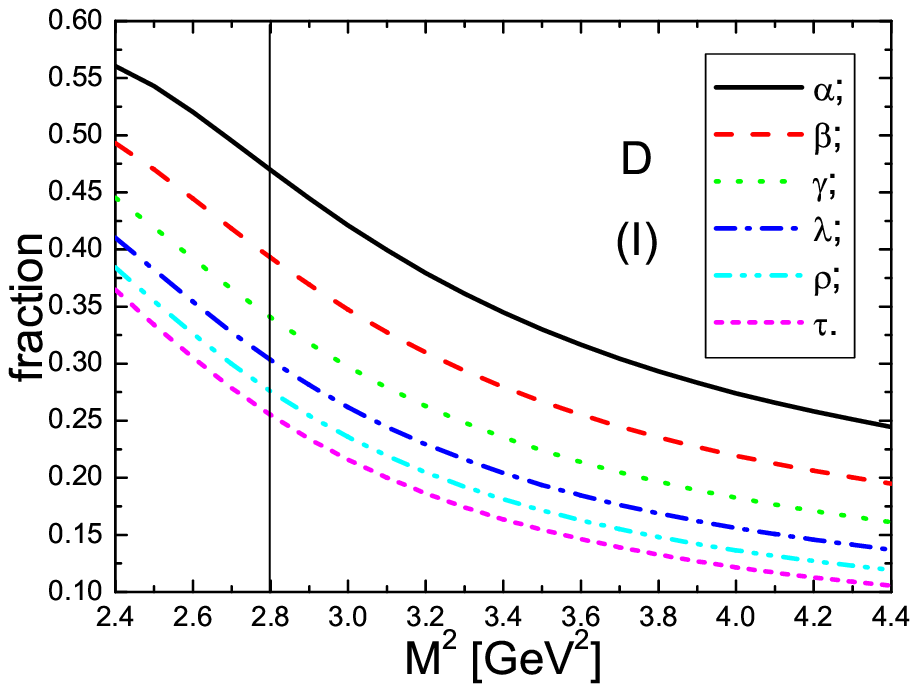}
 \includegraphics[totalheight=4.5cm,width=6cm]{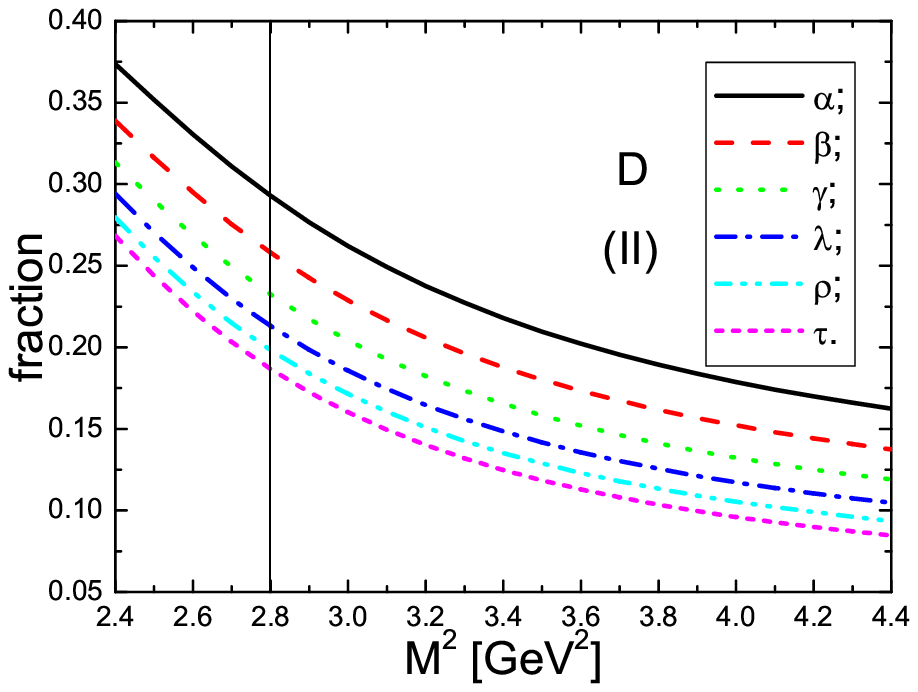}
     \caption{ The contributions from different terms with variation of the Borel
   parameter $M^2$  in the operator product expansion. The $A$,
   $B$, $C$ and $D$  correspond to the contributions from the $\langle \bar{q}g_s\sigma G q \rangle^2$ term,
   the $\langle \bar{q} q \rangle^2$ +$\langle \bar{q} q \rangle\langle \bar{q}g_s\sigma G q \rangle$ term,
   the $\langle \frac{\alpha_s GG}{\pi} \rangle $ term and the $\langle \frac{\alpha_s
GG}{\pi} \rangle $+$\langle \frac{\alpha_s GG}{\pi} \rangle
\left[\langle \bar{q} q \rangle +\langle \bar{q}g_s\sigma G q
\rangle+ \langle \bar{q} q \rangle^2\right]$ term, respectively. The
   (I) and (II)  denote the $c\bar{c}s\bar{s}$
  and $c\bar{c}q\bar{q}$ channels, respectively. The
notations
   $\alpha$, $\beta$, $\gamma$, $\lambda$, $\rho$ and $\tau$  correspond to the threshold
   parameters $s_0=23\,\rm{GeV}^2$,
   $24\,\rm{GeV}^2$, $25\,\rm{GeV}^2$, $26\,\rm{GeV}^2$, $27\,\rm{GeV}^2$ and $28\,\rm{GeV}^2$, respectively.}
\end{figure}

\begin{figure}
 \centering
 \includegraphics[totalheight=4.5cm,width=6cm]{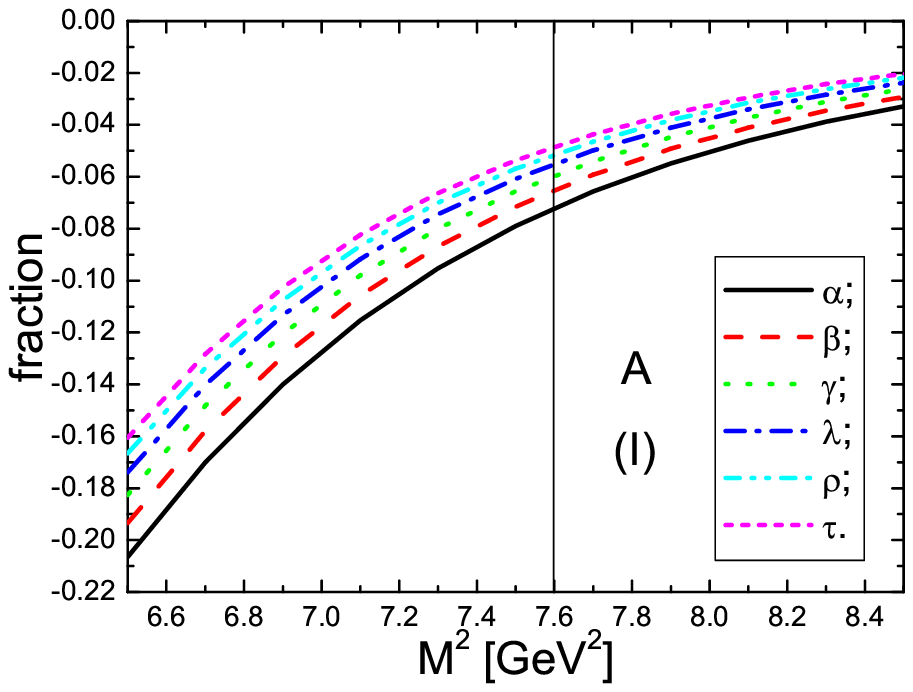}
 \includegraphics[totalheight=4.5cm,width=6cm]{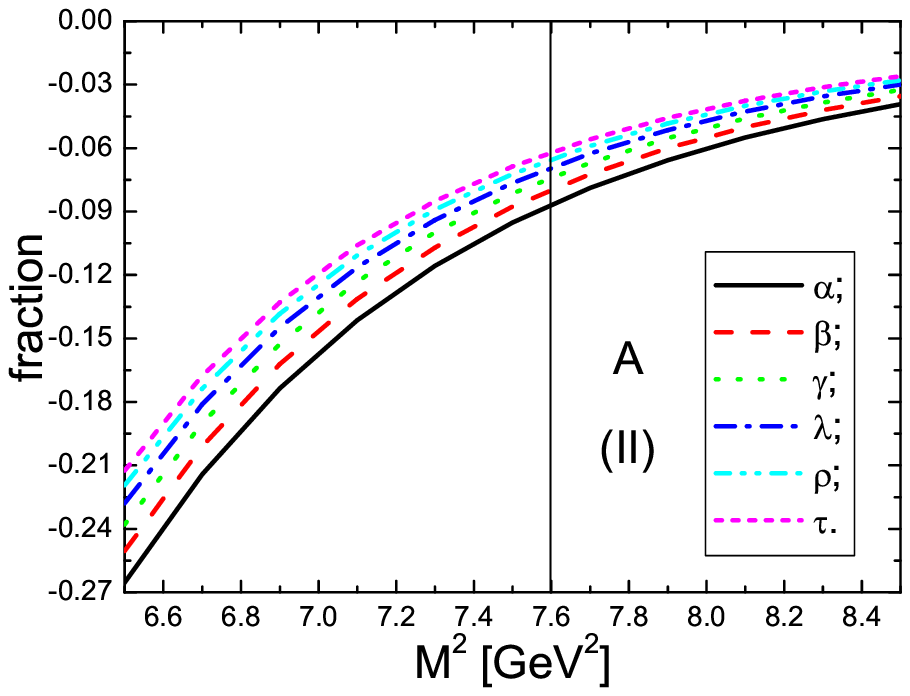}
 \includegraphics[totalheight=4.5cm,width=6cm]{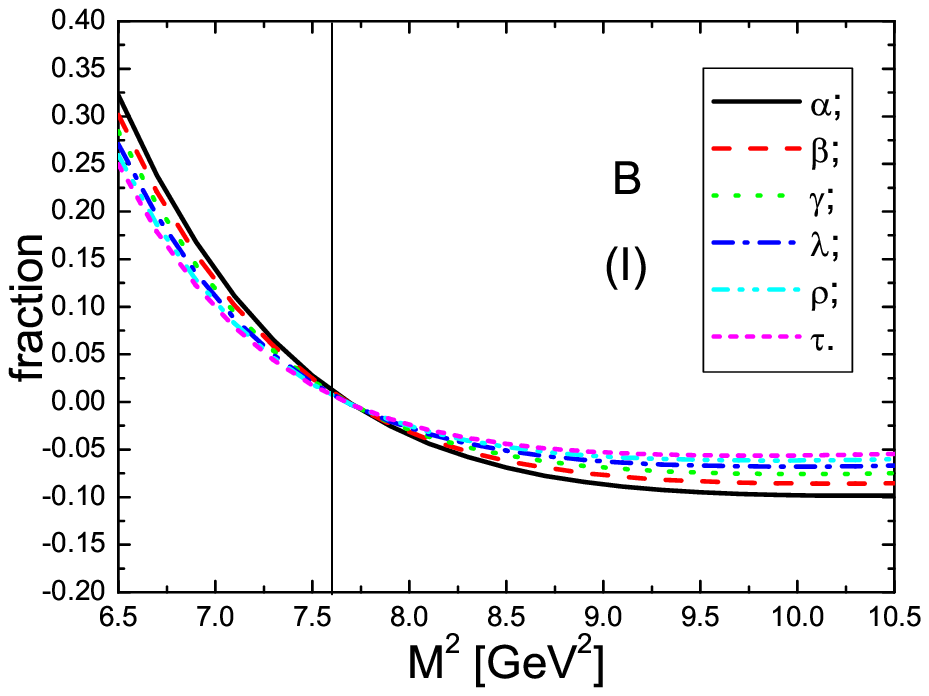}
 \includegraphics[totalheight=4.5cm,width=6cm]{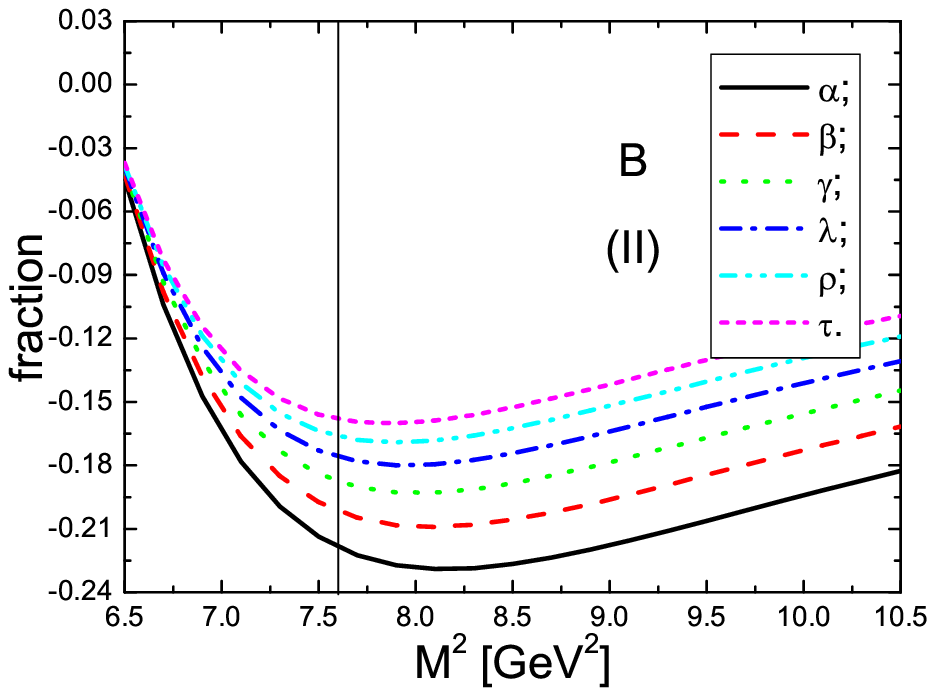}
 \includegraphics[totalheight=4.5cm,width=6cm]{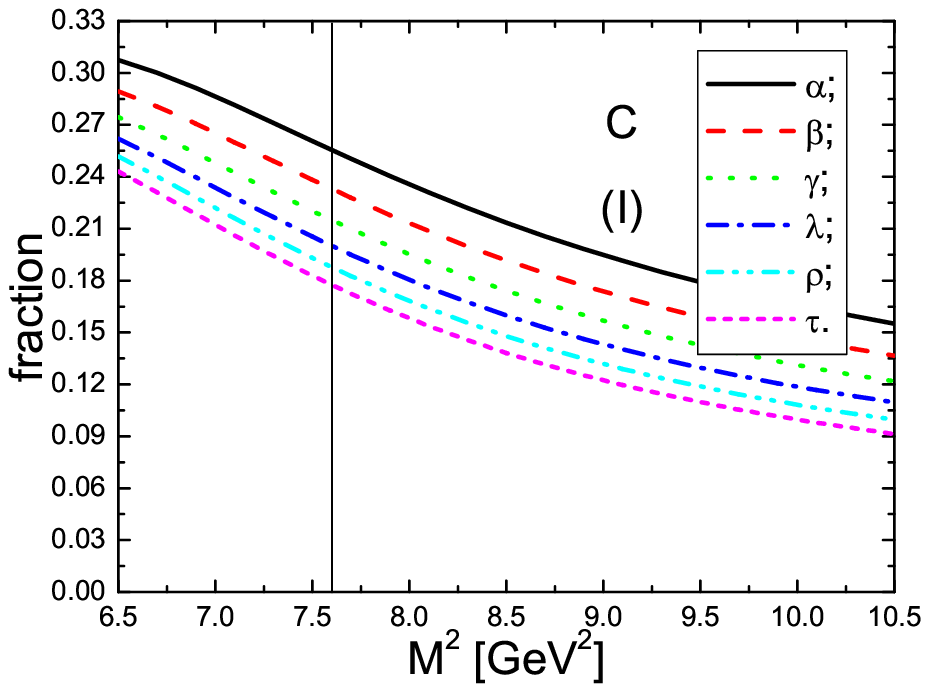}
 \includegraphics[totalheight=4.5cm,width=6cm]{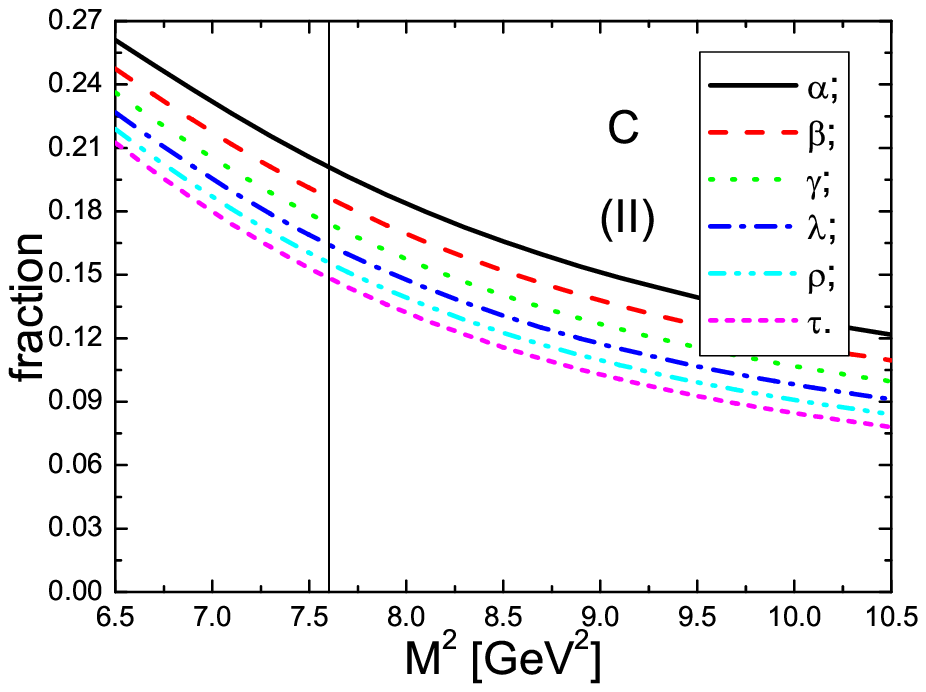}
 \includegraphics[totalheight=4.5cm,width=6cm]{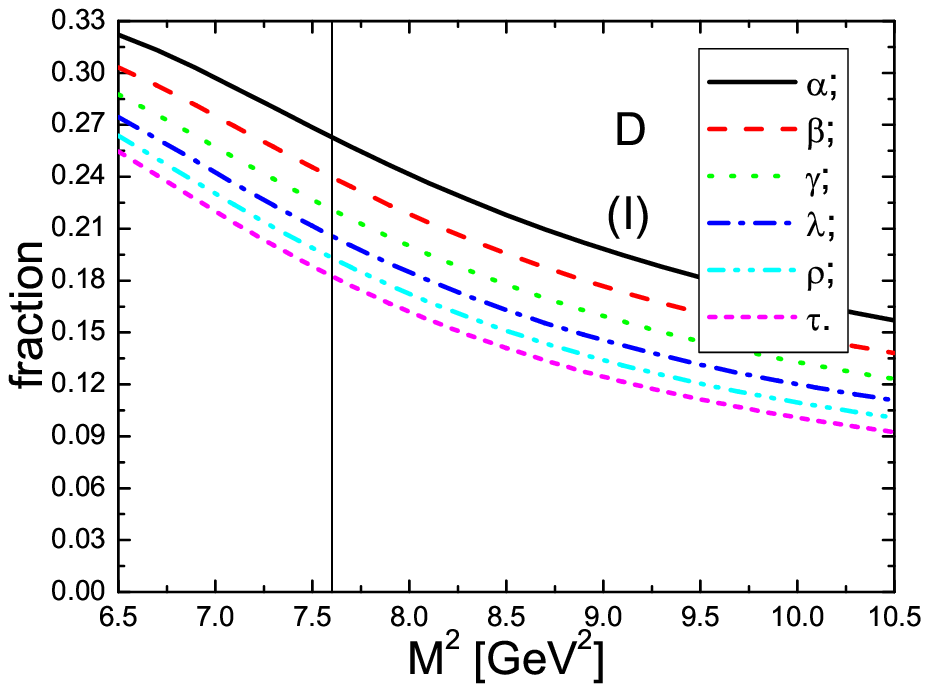}
 \includegraphics[totalheight=4.5cm,width=6cm]{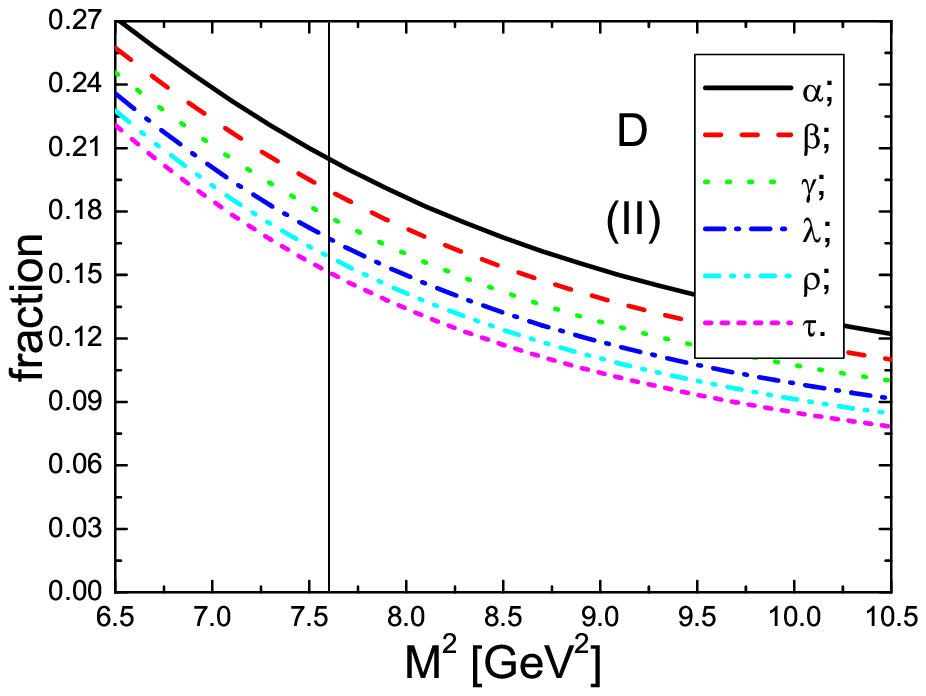}
     \caption{ The contributions from different terms with variation of the Borel
   parameter $M^2$  in the operator product expansion. The $A$,
   $B$, $C$ and $D$  correspond to the contributions from the $\langle \bar{q}g_s\sigma G q \rangle^2$ term,
   the $\langle \bar{q} q \rangle^2$ +$\langle \bar{q} q \rangle\langle \bar{q}g_s\sigma G q \rangle$ term,
   the $\langle \frac{\alpha_s GG}{\pi} \rangle $ term and the $\langle \frac{\alpha_s
GG}{\pi} \rangle $+$\langle \frac{\alpha_s GG}{\pi} \rangle
\left[\langle \bar{q} q \rangle +\langle \bar{q}g_s\sigma G q
\rangle+ \langle \bar{q} q \rangle^2\right]$ term, respectively. The
   (I) and (II)  denote the $b\bar{b}s\bar{s}$
  and $b\bar{b}q\bar{q}$ channels, respectively. The
notations
   $\alpha$, $\beta$, $\gamma$, $\lambda$, $\rho$ and $\tau$  correspond to the threshold
   parameters $s_0=142\,\rm{GeV}^2$,
   $144\,\rm{GeV}^2$, $146\,\rm{GeV}^2$, $148\,\rm{GeV}^2$, $150\,\rm{GeV}^2$ and $152\,\rm{GeV}^2$, respectively.}
\end{figure}

\begin{figure}
 \centering
 \includegraphics[totalheight=5cm,width=6cm]{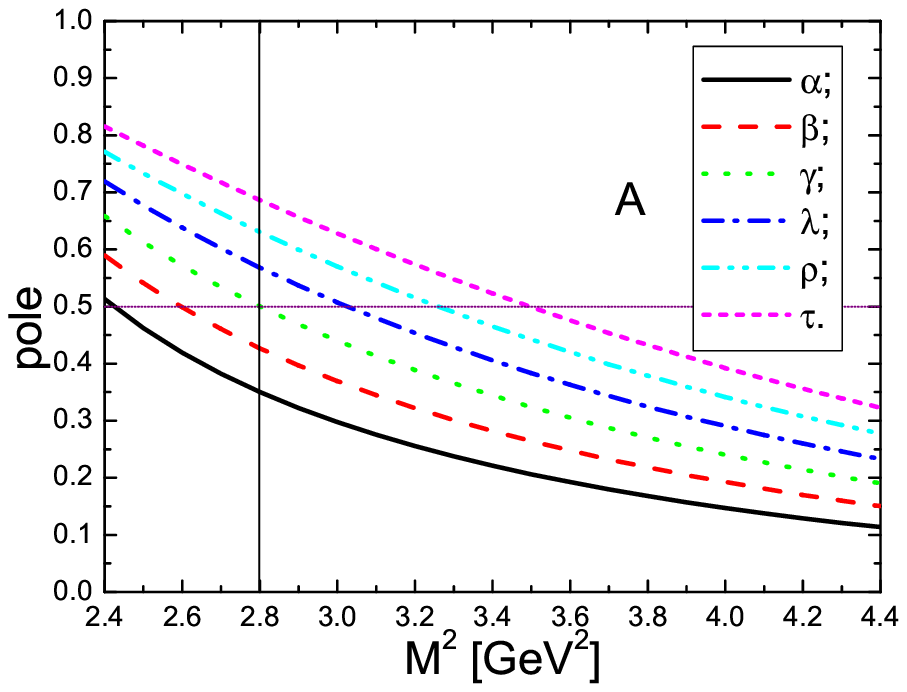}
\includegraphics[totalheight=5cm,width=6cm]{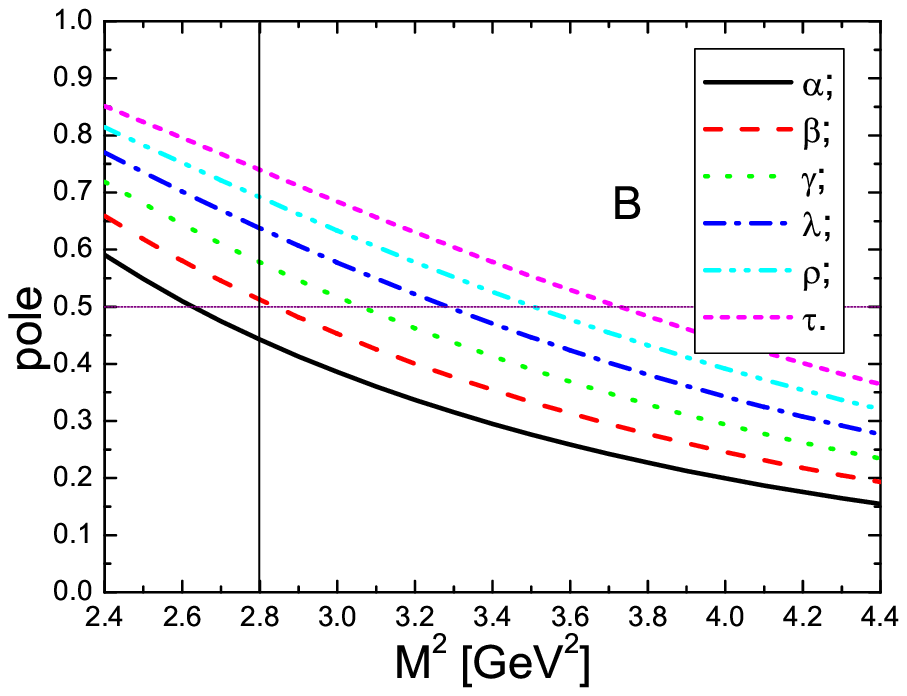}
\includegraphics[totalheight=5cm,width=6cm]{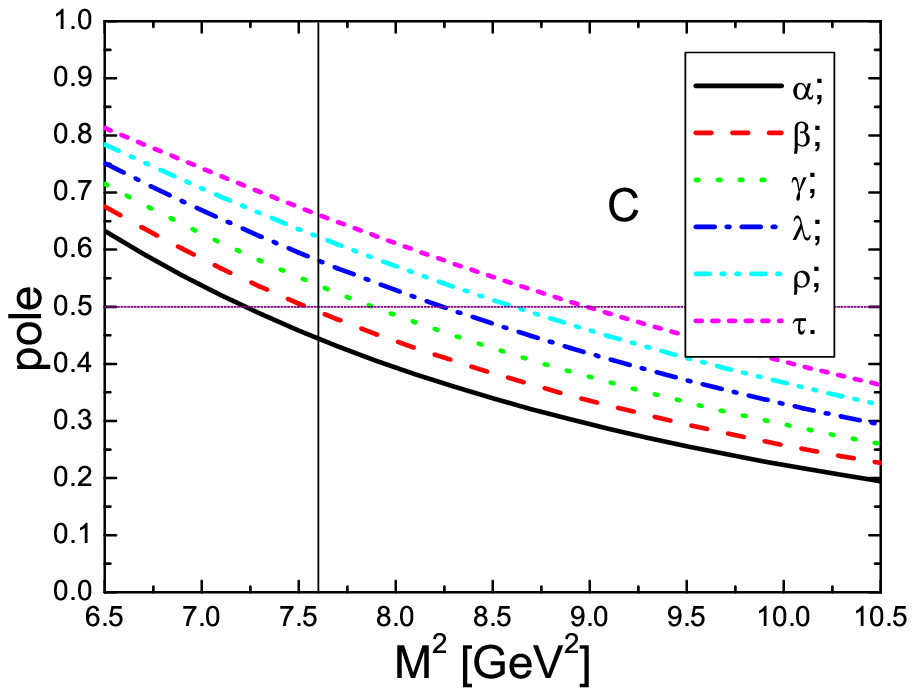}
\includegraphics[totalheight=5cm,width=6cm]{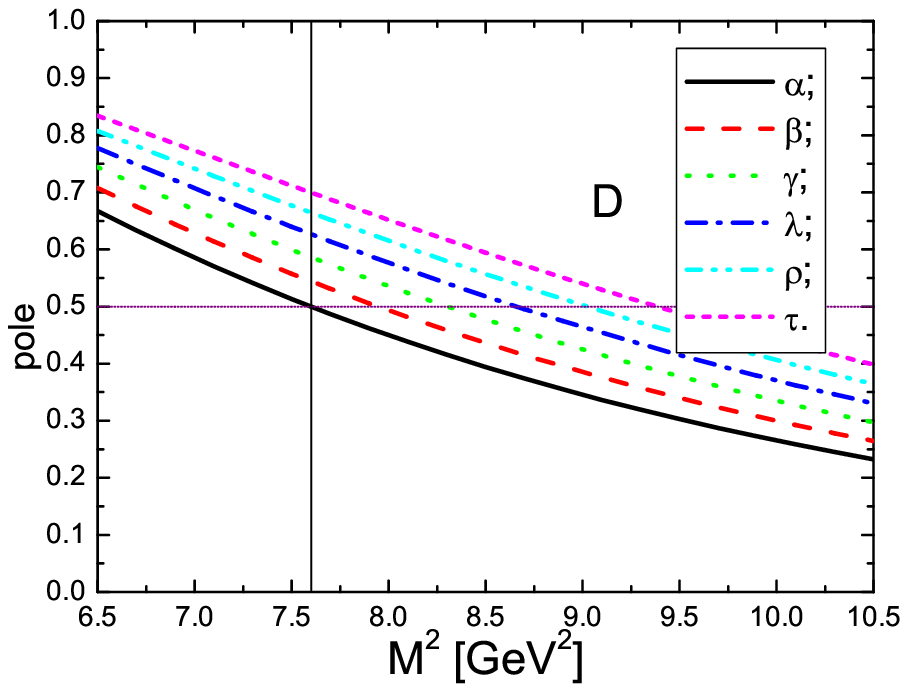}
   \caption{ The contributions from the pole terms with variation of the Borel parameter $M^2$. The $A$, $B$, $C$,
   and $D$ denote the $c\bar{c}s\bar{s}$,
    $c\bar{c}q\bar{q}$, $b\bar{b}s\bar{s}$
    and $b\bar{b}q\bar{q}$ channels, respectively.  In the hidden charm channels, the notations
   $\alpha$, $\beta$, $\gamma$, $\lambda$, $\rho$ and $\tau$  correspond to the threshold
   parameters $s_0=23\,\rm{GeV}^2$,
   $24\,\rm{GeV}^2$, $25\,\rm{GeV}^2$, $26\,\rm{GeV}^2$, $27\,\rm{GeV}^2$ and $28\,\rm{GeV}^2$, respectively
   ;  while in the hidden bottom channels they correspond to
    the threshold
   parameters  $s_0=142\,\rm{GeV}^2$,
   $144\,\rm{GeV}^2$, $146\,\rm{GeV}^2$, $148\,\rm{GeV}^2$, $150\,\rm{GeV}^2$ and $152\,\rm{GeV}^2$, respectively.}
\end{figure}

\begin{figure}
 \centering
  \includegraphics[totalheight=5cm,width=6cm]{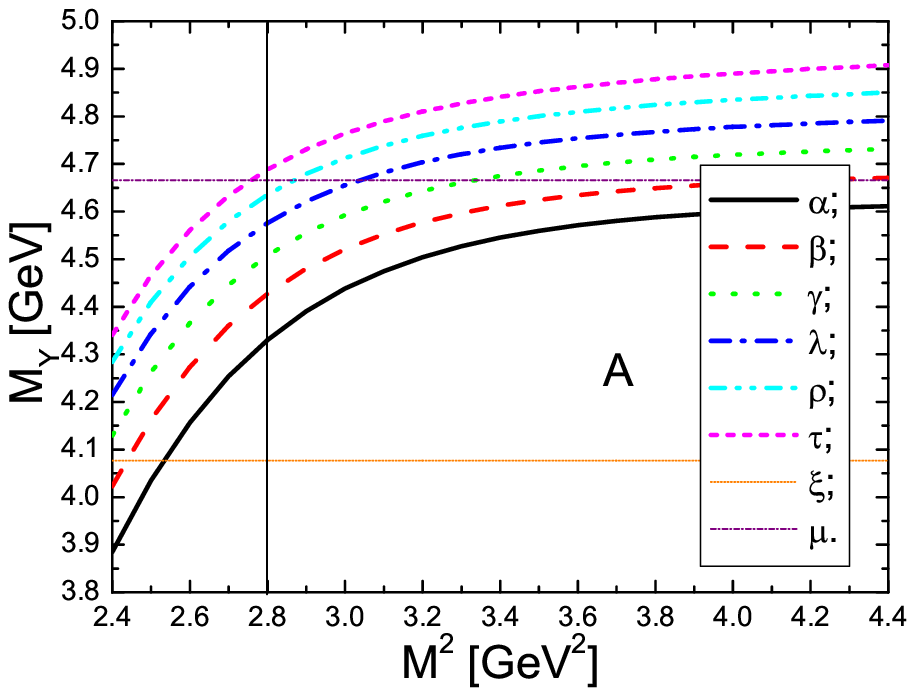}
  \includegraphics[totalheight=5cm,width=6cm]{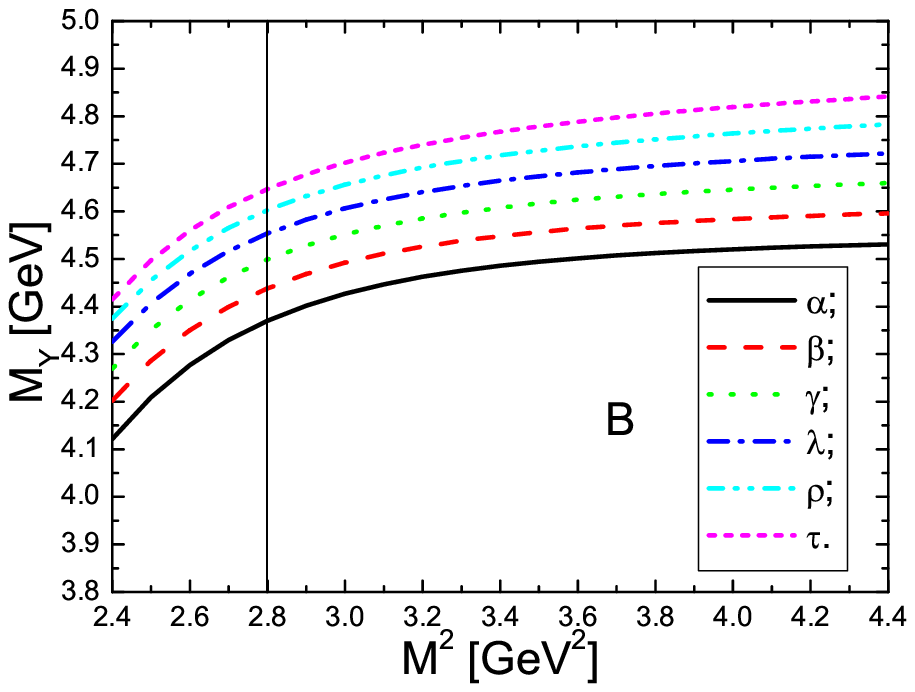}
  \includegraphics[totalheight=5cm,width=6cm]{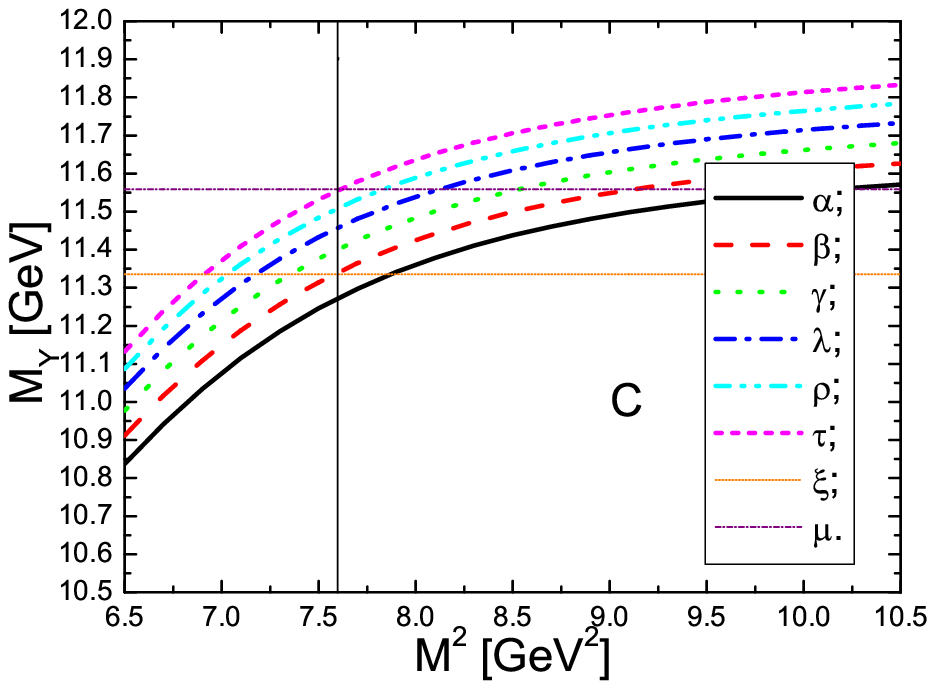}
  \includegraphics[totalheight=5cm,width=6cm]{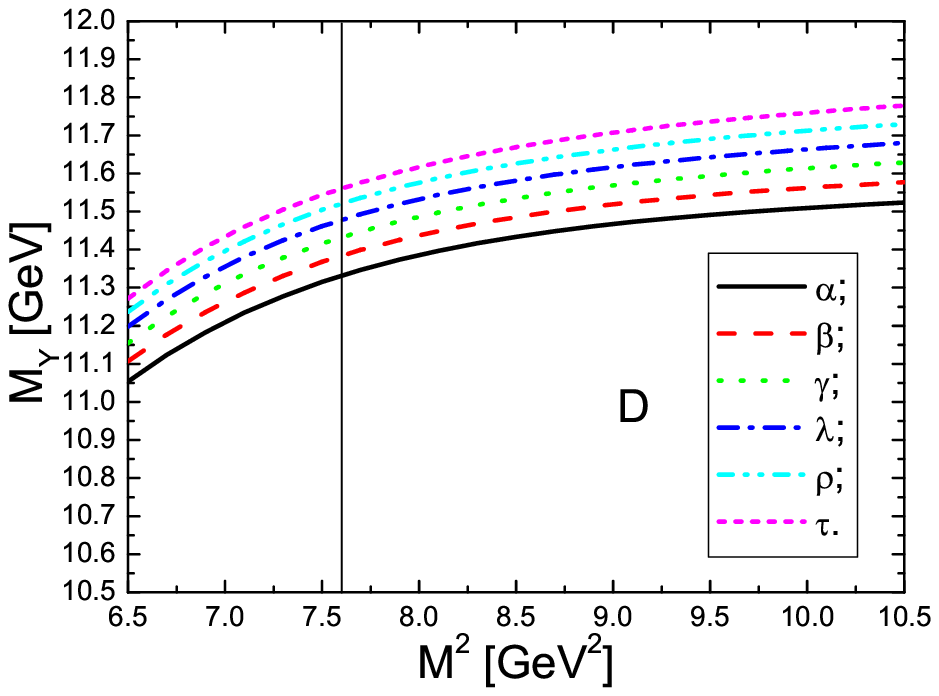}
   \caption{ The masses of the vector bound  states with variation of the Borel parameter $M^2$. The $A$, $B$, $C$,
   and $D$ denote the $c\bar{c}s\bar{s}$,
    $c\bar{c}q\bar{q}$, $b\bar{b}s\bar{s}$,
   and $b\bar{b}q\bar{q}$ channels, respectively. In the hidden charm channels, the notations
   $\alpha$, $\beta$, $\gamma$, $\lambda$, $\rho$ and $\tau$  correspond to the threshold
   parameters $s_0=23\,\rm{GeV}^2$,
   $24\,\rm{GeV}^2$, $25\,\rm{GeV}^2$, $26\,\rm{GeV}^2$, $27\,\rm{GeV}^2$ and $28\,\rm{GeV}^2$, respectively
   ;  while in the hidden bottom channels they correspond to
    the threshold
   parameters  $s_0=142\,\rm{GeV}^2$,
   $144\,\rm{GeV}^2$, $146\,\rm{GeV}^2$, $148\,\rm{GeV}^2$, $150\,\rm{GeV}^2$ and $152\,\rm{GeV}^2$, respectively. The $\xi$ and $\mu$
   denote the $J/\psi -f_0(980)$ and $\psi'-f_0(980)$ thresholds respectively in the $c\bar{c}s\bar{s}$ channel,
   while in the $b\bar{b}s\bar{s}$ channel they
   correspond to $\Upsilon''- f_0(980)$ and $\Upsilon'''- f_0(980)$ thresholds respectively.}
\end{figure}

\begin{figure}
 \centering
  \includegraphics[totalheight=5cm,width=6cm]{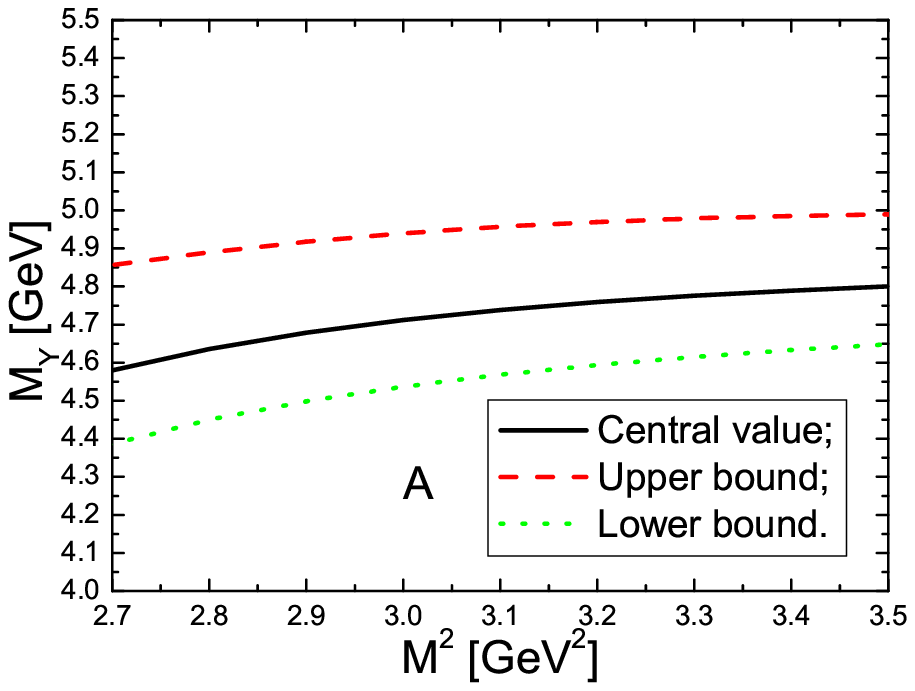}
  \includegraphics[totalheight=5cm,width=6cm]{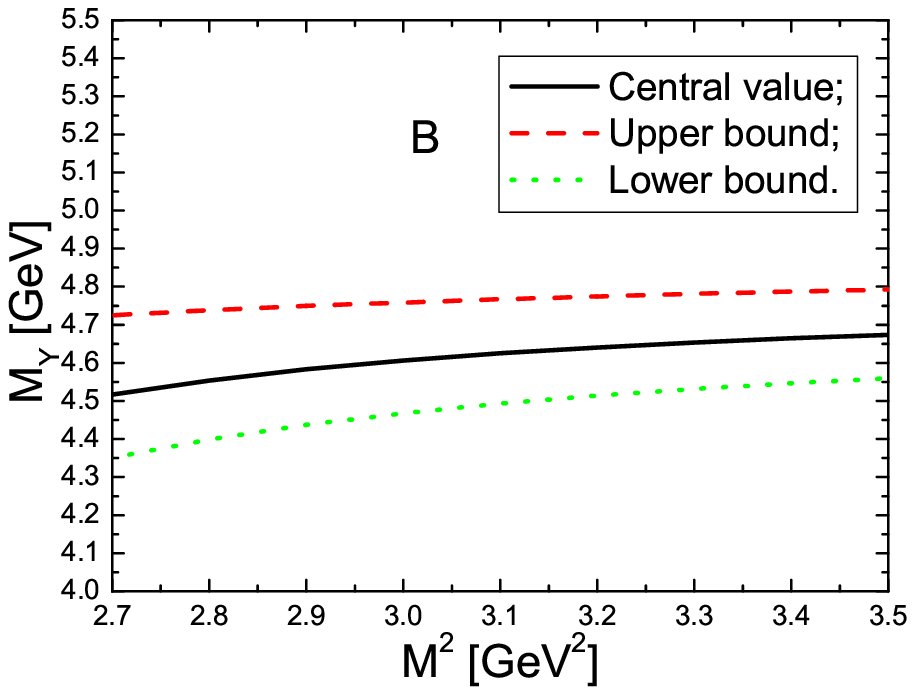}
  \includegraphics[totalheight=5cm,width=6cm]{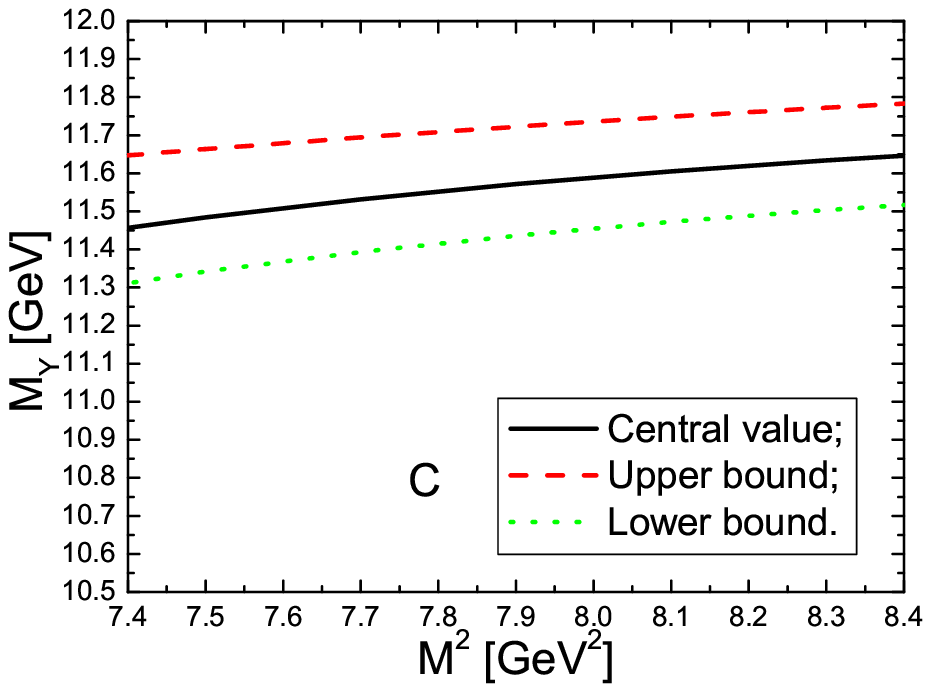}
  \includegraphics[totalheight=5cm,width=6cm]{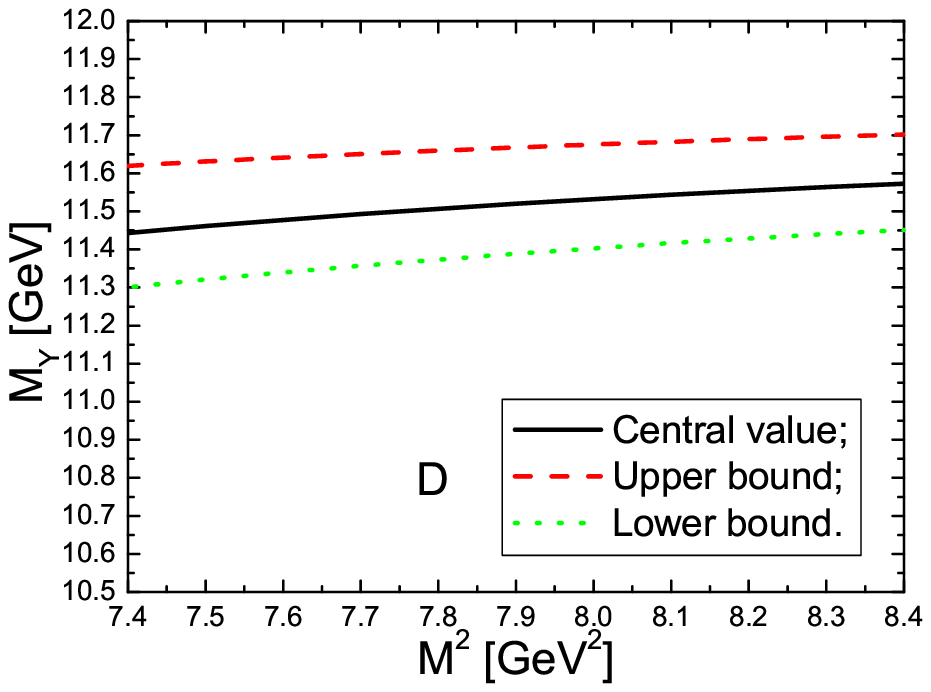}
   \caption{ The masses of the vector bound  states with variation of the Borel parameter $M^2$. The $A$, $B$, $C$,
   and $D$ denote the $c\bar{c}s\bar{s}$,
    $c\bar{c}q\bar{q}$, $b\bar{b}s\bar{s}$,
   and $b\bar{b}q\bar{q}$ channels, respectively. }
\end{figure}

\begin{figure}
 \centering
  \includegraphics[totalheight=5cm,width=6cm]{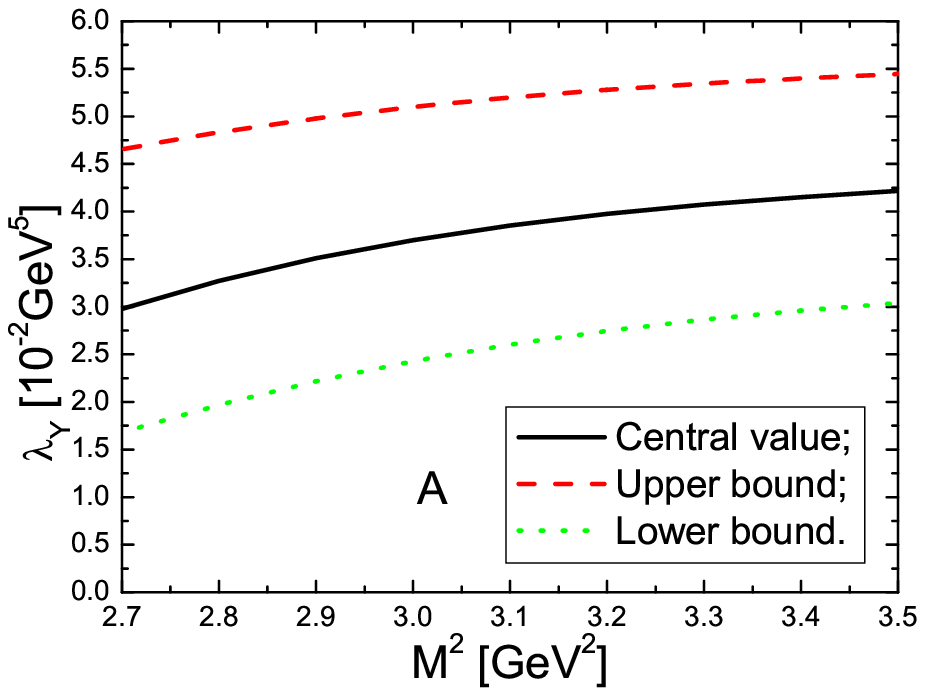}
  \includegraphics[totalheight=5cm,width=6cm]{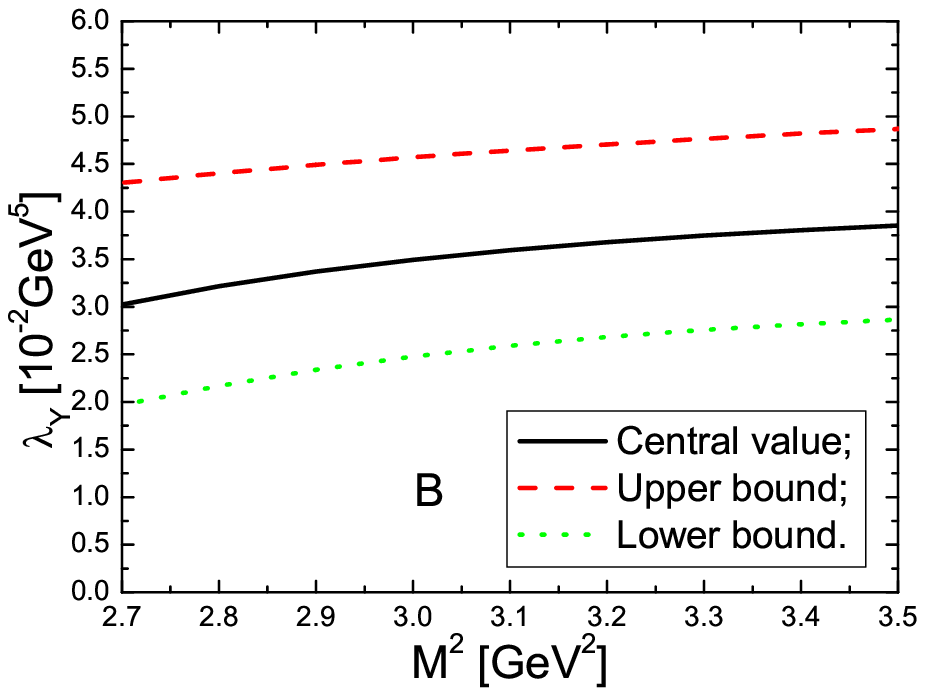}
  \includegraphics[totalheight=5cm,width=6cm]{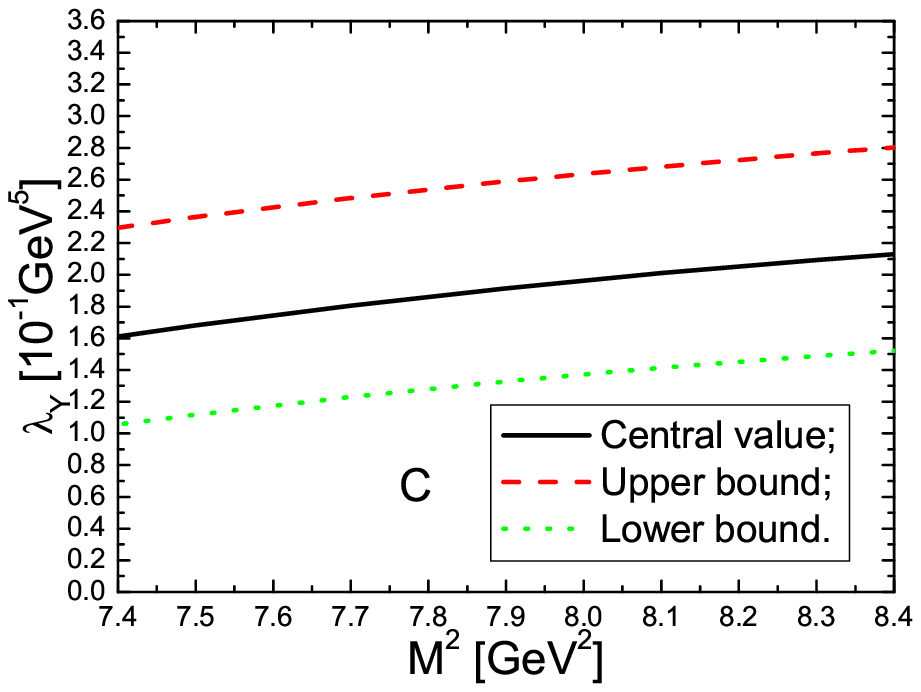}
  \includegraphics[totalheight=5cm,width=6cm]{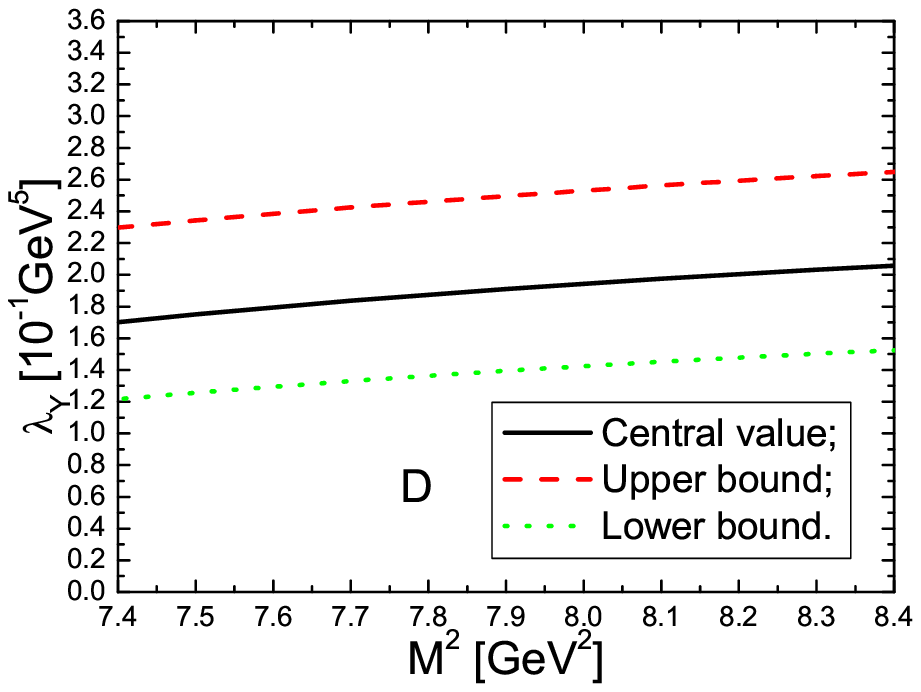}
    \caption{ The pole residues of the vector bound states with variation of the Borel parameter $M^2$. The $A$, $B$, $C$,
   and $D$ denote the $c\bar{c}s\bar{s}$,
    $c\bar{c}q\bar{q}$, $b\bar{b}s\bar{s}$,
   and $b\bar{b}q\bar{q}$ channels, respectively.}
\end{figure}

In Fig.3, we show the  contributions from the pole terms with
variation of the Borel parameters $M^2$ and the threshold parameters
$s_0$. If the pole dominance criterion is satisfied, the threshold
parameter $s_0$ increases   with  the Borel parameter $M^2$
monotonously. From
 Fig.3-A, we can see that  the pole dominance criterion cannot
be satisfied at the values  $s_0\leq 25 \,\rm{GeV}^2$ and $M^2\geq
2.8\,\rm{GeV}^2$ in the $c\bar{c}s\bar{s}$ channel,  the threshold
parameter $s_0$ has to  be pushed to larger value.

The pole contributions are larger than  $45\%$ at the values $M^2
\leq 3.2 \, \rm{GeV}^2 $ and $s_0\geq
25\,\rm{GeV}^2,\,26\,\rm{GeV}^2$ for the $c\bar{c}q\bar{q}$,
    $c\bar{c}s\bar{s}$
channels respectively; and larger than  $50\%$ at the values $M^2
\leq 8.2 \, \rm{GeV}^2 $, $s_0\geq
146\,\rm{GeV}^2,\,148\,\rm{GeV}^2$ for  the $b\bar{b}q\bar{q}$
  and $b\bar{b}s\bar{s}$ channels respectively. Again we
take the uniform Borel parameter $M^2_{max}$, i.e. $M^2_{max}\leq
3.2 \, \rm{GeV}^2$ and $M^2_{max}\leq 8.2 \, \rm{GeV}^2$ for the
hidden charm  and hidden bottom channels, respectively.

If we take  uniform pole contributions,   the interpolating current
with more $s$ quarks requires larger threshold parameter due to the
$SU(3)$ breaking effects, see Fig.3. The threshold parameters in the
$c\bar{c}q\bar{q}$ and $b\bar{b}q\bar{q}$ channels are slightly
smaller than the ones in the $c\bar{c}s\bar{s}$ and
$b\bar{b}s\bar{s}$ channels respectively. In this article, the
threshold parameters are taken as $s_0=(26\pm1)\,\rm{GeV}^2$,
$(27\pm1)\,\rm{GeV}^2$, $(148\pm2)\,\rm{GeV}^2$ and
$(150\pm2)\,\rm{GeV}^2$ for the $c\bar{c}q\bar{q}$,
    $c\bar{c}s\bar{s}$, $b\bar{b}q\bar{q}$
     and $b\bar{b}s\bar{s}$ channels, respectively;
   the Borel parameters are taken as $M^2=(2.8-3.2)\,\rm{GeV}^2$ and
   $(7.6-8.2)\,\rm{GeV}^2$ for the
hidden charm and hidden bottom channels, respectively. In those
regions, the pole contributions are about  $(45-69)\%$, $(46-69)\%$,
$(50-66)\%$ and $(51-67)\%$ for  the $c\bar{c}s\bar{s}$,
    $c\bar{c}q\bar{q}$, $b\bar{b}s\bar{s}$
     and $b\bar{b}q\bar{q}$ channels, respectively;  the two criteria of the QCD sum rules
are fully  satisfied  \cite{SVZ79,Reinders85}. Naively, we expect
the bound state with the scalar meson $f_0(980)$ will have larger
mass than the corresponding one with the scalar meson
$\sigma(400-1200)$, our numerical calculations confirm this
conjecture, see Fig.4. Although smaller threshold parameters lead to
slower convergent behavior in the operator product expansion, the
two criteria of the QCD sum rules are still satisfied, one can
consult Figs.1-2.

The Borel windows $M_{max}^2-M_{min}^2$ change with  variations of
the  threshold parameters $s_0$, see Fig.3. In this article, the
Borel windows  are taken as  $0.4\,\rm{GeV}^2$ and $0.6\,\rm{GeV}^2$
for the hidden charm and hidden bottom channels respectively, they
are small enough. Furthermore, we take uniform Borel windows and
smear the dependence on the threshold parameters $s_0$.  If we take
larger threshold parameters,  the Borel windows are larger and the
resulting  masses are larger, see Fig.4. In this article, we intend
calculate the possibly  lowest  masses which are supposed to be the
ground state masses  by imposing the two criteria of the QCD sum
rules.

In Fig.4, we plot the  bound state masses $M_Y$ with variation of
the Borel parameters and the threshold parameters.  The hidden charm
current $\bar{c}(x)\gamma_\mu c(x)$ can interpolate the charmonia
$J/\psi$, $\psi'$, $\psi(3770)$, $\psi(4040)$, $\psi(4160)$,
$\psi(4415)$, $\cdots$; while the hidden bottom current
$\bar{b}(x)\gamma_\mu b(x)$ can interpolate the bottomonia
$\Upsilon$, $\Upsilon'$, $\Upsilon''$, $\Upsilon'''$,
$\Upsilon''''$, $\cdots$ \cite{PDG}. The currents $J_\mu(x)$  have
non-vanishing couplings with the bound states $J/\psi f_0(980)$,
$\psi'f_0(980)$, $\psi''f_0(980)$, $\cdots$ and $\Upsilon f_0(980)$,
$\Upsilon'f_0(980)$, $\Upsilon''f_0(980)$, $\Upsilon'''f_0(980)$,
$\cdots$, respectively. From Figs.3-A,3-C,4-A,4-C, we can see that
the QCD sum rules support existence of the $\psi' f_0(980)$ and
$\Upsilon'''f_0(980)$ bound states, the nominal thresholds  of the
$J/\psi-f_0(980)$ and $\Upsilon''-f_0(980)$ systems are too low, and
we cannot reproduce the $J/\psi f_0(980)$ and $\Upsilon''f_0(980)$
bound states. Our numerical results support the conjecture of
Voloshin et al, i.e. a formation of hadro-charmonium is favored for
higher charmonium resonances $\psi'$ and $\chi_{cJ}$ as compared to
the lowest states $J/\psi$ and $\eta_c$ \cite{Voloshin0803}.

In this article, we  intend  prove that the $\psi' f_0(980)$ and
$\Upsilon'''f_0(980)$ bound states can be reproduced by the QCD sum
rules, the charmonium-like state $Y(4660)$ has the possibility to be
a $\psi' f_0(980)$ bound  state.

Taking into account all uncertainties of the input parameters,
finally we obtain the values of the masses and pole resides of
 the  vector bound states  $Y$, which are  shown in Figs.5-6 and Tables 1-2.
In this article,  we calculate the uncertainties $\delta$  with the
formula
\begin{eqnarray}
\delta=\sqrt{\sum_i\left(\frac{\partial f}{\partial
x_i}\right)^2\mid_{x_i=\bar{x}_i} (x_i-\bar{x}_i)^2}\,  ,
\end{eqnarray}
 where the $f$ denote  the
hadron mass  $M_Y$ and the pole residue $\lambda_Y$,  the $x_i$
denote the input QCD parameters $m_c$, $m_b$, $\langle \bar{q}q
\rangle$, $\langle \bar{s}s \rangle$, $\cdots$. As the partial
 derivatives   $\frac{\partial f}{\partial x_i}$ are difficult to carry
out analytically, we take the  approximation $\left(\frac{\partial
f}{\partial x_i}\right)^2 (x_i-\bar{x}_i)^2\approx
\left[f(\bar{x}_i\pm \Delta x_i)-f(\bar{x}_i)\right]^2$ in the
numerical calculations.

From Tables 1-2, we can see that the uncertainties of the masses
$M_Y$ are rather small (about $5\%$ in the hidden charm channels and
$2\%$ in the hidden bottom channels),  while the uncertainties of
the pole residues $\lambda_{Y}$ are rather large (about
$(30-50)\%$). The uncertainties of the input parameters ($\langle
\bar{q}q \rangle$, $\langle \bar{s}s \rangle$, $\langle
\bar{s}g_s\sigma G s \rangle$, $\langle \bar{q}g_s\sigma G q
\rangle$,
 $m_s$,  $m_c$ and $m_b$) vary in the range
$(2-25)\%$, the uncertainties of the pole  residues $\lambda_{Y}$
are reasonable. We obtain the  squared masses   $M_Y^2$ through a
fraction, see Eq.(7), the uncertainties in the numerator and
denominator which origin from a given input parameter (for example,
$\langle \bar{s}s \rangle$, $\langle \bar{s}g_s\sigma G s \rangle$)
cancel out with each other, and result in small net uncertainty.

\begin{table}
\begin{center}
\begin{tabular}{|c|c|c|c|c|}
\hline\hline bound states & $M_Y$ ($ \rm{GeV}$) & $M_{\psi'/\Upsilon'''}+M_{f_0/\sigma}$ ($ \rm{GeV}$)&$M_Y$ ($ \rm{GeV}$)$*$\\
\hline
      $c\bar{c}s\bar{s}$  &$4.71\pm0.26$ &$4.666$& $4.63$\\ \hline
           $c\bar{c}q\bar{q} $ &$4.59\pm0.19$& $4.086-4.886$& $4.56$\\      \hline
    $b\bar{b}s\bar{s}$  &$11.57\pm0.20$ &$11.559$&$11.56$\\ \hline
            $ b\bar{b}q\bar{q} $ &$11.52\pm0.18$ &$10.979-11.779$&$11.51$\\      \hline
    \hline
\end{tabular}
\end{center}
\caption{ The masses  of the bound  states, we use the star $*$ to
denote  the central values from the sum rules where the perturbative
contributions are multiplied by a factor 2. }
\end{table}

\begin{table}
\begin{center}
\begin{tabular}{|c|c|c|c|}
\hline\hline bound states & $\lambda_{Y}$ ($10^{-2} \rm{GeV}^5$)& $\lambda_{Y}$ ($10^{-2} \rm{GeV}^5$)$*$\\
\hline
      $c\bar{c}s\bar{s}$  &$3.70^{+1.58}_{-1.74}$&$5.23$\\ \hline
            $c\bar{c}q\bar{q} $ & $3.49^{+1.21}_{-1.32}$&$4.84$\\      \hline
    $b\bar{b}s\bar{s}$   &$19.2\pm8.2$&$28.6$\\ \hline
            $ b\bar{b}q\bar{q} $  &$19.2\pm6.7$&$27.3$\\      \hline
    \hline
\end{tabular}
\end{center}
\caption{ The  pole residues of the bound  states, we use the star
$*$ to denote  the central values from the sum rules where the
perturbative contributions are multiplied by a factor 2. }
\end{table}

In table 1, we also present the  nominal thresholds of the
$\psi'-f_0(980)$, $\psi'-\sigma(400-1200)$,  $\Upsilon'''-f_0(980)$
and $\Upsilon'''-\sigma(400-1200)$ systems. From the table, we can
see that  the $Y(4660)$ can be tentatively identified as the
$\psi'f_0(980)$ bound state. The predicted mass of the
$\psi'\sigma(400-1200)$ bound state is about
$(4.59\pm0.19)\rm{GeV}$, while the nominal threshold of the
$\psi'-\sigma(400-1200)$ system is about $(4.086-4.886)\, \rm{GeV}$.
There maybe exist such a bound state. The $\psi'\sigma(400-1200)$
bound state can  be produced in the initial state radiation  process
$e^+e^-\to\gamma_{ISR}\pi^+\pi^-\psi'$ or in the exclusive decays of
the $B$ meson through $b\rightarrow c\bar{c}q$ at the quark level.
There still lack experimental candidates to identify the
$\psi'\sigma(400-1200)$ bound state, such   a bound state is
difficult to observe  due to the broad width of the scalar meson
$\sigma(400-1200)$.

In the $b\bar{b}s\bar{s}$ channel, the numerical result
$M_Y=11.57\pm0.20\,\rm{GeV}$ indicates that there maybe exist a
$\Upsilon'''f_0(980)$ bound state, which is consistent with the
nominal threshold $M_{\Upsilon'''}+M_{f_0}=11.559\,\rm{GeV}$, while
the  nominal thresholds $M_{\Upsilon}+M_{f_0}=10.44\,\rm{GeV}$,
$M_{\Upsilon'}+M_{f_0}=11.00\,\rm{GeV}$,
$M_{\Upsilon''}+M_{f_0}=11.335\,\rm{GeV}$ are too low. The scalar
meson $\sigma(400-1200)$ is rather broad with the Breit-Wigner mass
formula $(400-1200)-i(250-500)$ \cite{PDG}. Considering the $SU(3)$
symmetry of the light flavor quarks, we can obtain the conclusion
tentatively that there maybe exist the $\psi'\sigma(400-1200)$ and
$\Upsilon'''\sigma(400-1200)$ bound states which lie in the regions
$(4.086-4.886) \,\rm{GeV}$ and $(10.979-11.779)\,\rm{GeV}$,
respectively. As the energy gaps between the $\Upsilon$'s are rather
small and the scalar meson $\sigma(400-1200)$ is broad enough, there
maybe   exist the $\Upsilon\sigma(400-1200)$,
$\Upsilon'\sigma(400-1200)$ and $\Upsilon''\sigma(400-1200)$ bound
states. We cannot draw  decisive conclusion with the QCD sum rules
alone.

At the energy scale $\mu=1\, \rm{GeV}$, $\frac{\alpha_s}{\pi}\approx
0.19$ \cite{Davier2006}, if the perturbative $\mathcal
{O}(\alpha_s)$ corrections to the perturbative term are companied
with large numerical factors, $1+\xi(s,m_Q)\frac{\alpha_s}{\pi}$,
for example, $\xi(s,m_Q) >\frac{\pi}{\alpha_s}\approx 5$, the
contributions may be large. We can make a crude estimation by
multiplying the perturbative term  with a numerical factor, say
$1+\xi(s,m_Q)\frac{\alpha_s}{\pi}=2$, the masses $M_Y$ decrease
slightly while  the  pole residues $\lambda_Y$ increase remarkably,
see Tables 1-2. The  main contribution comes from the perturbative
term, the large corrections in the numerator and denominator  cancel
out with each other (see Eq.(7)). In fact, the $\xi(s,m_Q)$ are
complicated functions of the energy $s$ and the mass $m_Q$, such a
crude estimation maybe underestimate the $\mathcal {O}(\alpha_s)$
corrections, the uncertainties originate from the $\mathcal
{O}(\alpha_s)$ corrections maybe larger.

The charmonia $J/\psi$, $\psi'$, $\psi(3770)$, $\psi(4040)$,
$\psi(4160)$, $\psi(4415)$, $\cdots$ and the bottomonia $\Upsilon$,
$\Upsilon'$, $\Upsilon''$, $\Upsilon'''$, $\Upsilon''''$, $\cdots$
 also have Fock states with additional $q\bar{q}$ components beside
 the $Q\bar{Q}$ components. The currents $J_\mu(x)$ and
 $\eta_\mu(x)$ may have non-vanishing couplings  with the charmonia
 and bottomonia, those couplings are supposed to be small, as the
 main Fock states of the charmonia
 and bottomonia are the $Q\bar{Q}$ components, and the  charmonia
 and bottomonia have much smaller masses than the corresponding
 molecular states $Y$.  

In this article, we take the assumption that the scalar mesons
$f_0(980)$ and $\sigma(400-1200)$ are the conventional $q\bar{q}$
mesons, or more precise, they have large $q\bar{q}$ components.
There are  hot controversies  about their nature, for example, the
conventional $q\bar{q}$ states (strongly affected  by the nearby
thresholds), the tetraquark states, the molecular states
\cite{Close2002,ReviewScalar}. In Ref.\cite{Wang2004}, we take the
scalar mesons $a_0(980)$ and $f_0(980)$ as the conventional
$q\bar{q}$ mesons,  study the strong couplings to the nearby
thresholds, and observe that the strong couplings are rather large.
Then we draw the conclusion that the $a_0(980)$ and $f_0(980)$ may
have a small $q\bar{q} $ kernel of the typical $q\bar{q}$ meson
size, strong coupling to the nearby $\bar{K}K$ threshold may result
in some tetraquark components (irrespective of a nucleon-like bound
state and a deuteron-like bound state) \cite{Wang2004}. The decay
$f_0(980)/\sigma(400-1200)\to
 \pi\pi,K\bar{K}$ can occur through the tetraquark quark components naturally.

The LHCb is a dedicated $b$ and $c$-physics precision experiment at
the LHC (large hadron collider). The LHC will be the world's most
copious  source of the $b$ hadrons, and  a complete spectrum of the
$b$ hadrons will be available through gluon fusion. In proton-proton
collisions at $\sqrt{s}=14\,\rm{TeV}$, the $b\bar{b}$ cross section
is expected to be $\sim 500\mu b$ producing $10^{12}$ $b\bar{b}$
pairs in a standard  year of running at the LHCb operational
luminosity of $2\times10^{32} \rm{cm}^{-2} \rm{sec}^{-1}$
\cite{LHC}. The bound  states $\Upsilon'''f_0(980)$ and
$\Upsilon'''\sigma(400-1200)$ predicted in the present work may be
observed at the LHCb, if they exist indeed. We can search for those
bound states in the $\Upsilon\pi\pi$, $\Upsilon'\pi\pi$,
$\Upsilon''\pi\pi$, $\Upsilon'''\pi\pi$, $\Upsilon K\bar{K}$,
$\Upsilon'K\bar{K}$, $\Upsilon''K\bar{K}$, $\Upsilon'''K\bar{K}$,
$\cdots$ invariant mass distributions.

\section{Conclusion}
In this article, we take the the vector charmonium-like state
$Y(4660)$ as the $\psi'f_0(980)$ bound state (irrespective of the
hadro-charmonium and the molecular state) tentatively, study its
mass  using the QCD sum rules, the numerical result
$M_Y=4.71\pm0.26\,\rm{GeV}$ is consistent with the experimental data
$4664\pm 11\pm 5~\rm{MeV}$. Considering the $SU(3)$ symmetry of the
light flavor quarks and the heavy quark symmetry, we also study the
bound states $\psi'\sigma(400-1200)$, $\Upsilon'''f_0(980)$ and
$\Upsilon'''\sigma(400-1200)$ with the QCD sum rules, and make
reasonable  predictions for their masses. Our predictions depend
heavily on  the two criteria (pole dominance and convergence of the
operator product expansion) of the QCD sum rules. We can search for
those bound states at the LHCb, the KEK-B or the Fermi-lab Tevatron.

\section*{Appendix}
The spectral densities at the level of the quark-gluon degrees of
freedom:
\begin{eqnarray}
 \rho_0(s)&=&\frac{3}{4096 \pi^6}
\int_{\alpha_{i}}^{\alpha_{f}}d\alpha \int_{\beta_{i}}^{1-\alpha}
d\beta\alpha\beta(1-\alpha-\beta)^2(s-\widetilde{m}^2_Q)^3(5s-\widetilde{m}^2_Q)\nonumber \\
&&+\frac{3m_Q^2}{1024 \pi^6} \int_{\alpha_{i}}^{\alpha_{f}}d\alpha
\int_{\beta_{i}}^{1-\alpha} d\beta
 (1-\alpha-\beta)^2(s-\widetilde{m}^2_Q)^3\,,
\end{eqnarray}

\begin{eqnarray}
\rho_{\langle
\bar{s}s\rangle}(s)&=&\frac{9m_s\langle\bar{s}s\rangle}{128 \pi^4}
\int_{\alpha_{i}}^{\alpha_{f}}d\alpha \int_{\beta_{i}}^{1-\alpha}
d\beta\alpha\beta(s-\widetilde{m}^2_Q)(3s-\widetilde{m}^2_Q)\nonumber \\
&&+\frac{9m_sm_Q^2\langle\bar{s}s\rangle}{64 \pi^4}
\int_{\alpha_{i}}^{\alpha_{f}}d\alpha \int_{\beta_{i}}^{1-\alpha}
d\beta (s-\widetilde{m}^2_Q) - \nonumber \\
&&\frac{m_s\langle\bar{s}g_s \sigma Gs\rangle}{32 \pi^4}
\int_{\alpha_{i}}^{\alpha_{f}}d\alpha \alpha(1-\alpha)
(2s-\widetilde{\widetilde{m}}^2_Q) -\frac{m_sm_Q^2\langle\bar{s}g_s
\sigma Gs\rangle}{32 \pi^4} \int_{\alpha_{i}}^{\alpha_{f}}d\alpha\,,
\end{eqnarray}

\begin{eqnarray}
\rho_{\langle \bar{s}s\rangle^2}(s)&=&-\frac{ \langle\bar{s}
s\rangle^2}{16 \pi^2} \int_{\alpha_{i}}^{\alpha_{f}}d\alpha
\alpha(1-\alpha)  (2s-\widetilde{\widetilde{m}}^2_Q) -\frac{
m_Q^2\langle\bar{s} s\rangle^2}{16 \pi^2}
\int_{\alpha_{i}}^{\alpha_{f}}d\alpha \nonumber \\
&& +\frac{ \langle\bar{s} s\rangle \langle\bar{s}g_s \sigma
Gs\rangle}{32 \pi^2} \int_{\alpha_{i}}^{\alpha_{f}}d\alpha
\alpha(1-\alpha)
\left[3+\left(3s+\frac{s^2}{M^2}\right)\delta(s-\widetilde{\widetilde{m}}^2_Q)
\right] \nonumber  \\
&& +\frac{ m_Q^2\langle\bar{s} s\rangle \langle\bar{s}g_s \sigma
Gs\rangle}{32 \pi^2} \int_{\alpha_{i}}^{\alpha_{f}}d\alpha
\left[1+\frac{s}{M^2}\right] \delta(s-\widetilde{\widetilde{m}}^2_Q)\nonumber  \\
&& +\frac{  \langle\bar{s}g_s \sigma Gs\rangle^2}{128 \pi^2M^2}
\int_{\alpha_{i}}^{\alpha_{f}}d\alpha \alpha(1-\alpha)s
\left[1+\frac{s}{M^2}+\frac{s^2}{2M^4}\right] \delta(s-\widetilde{\widetilde{m}}^2_Q)\nonumber  \\
&& +\frac{ 3m_Q^2 \langle\bar{s}g_s \sigma Gs\rangle^2}{768
\pi^2M^6} \int_{\alpha_{i}}^{\alpha_{f}}d\alpha s^2
\delta(s-\widetilde{\widetilde{m}}^2_Q)\, ,
\end{eqnarray}

\begin{eqnarray}
\rho^A_{\langle GG\rangle}(s)&=& \frac{3}{1024 \pi^4}
\int_{\alpha_{i}}^{\alpha_{f}}d\alpha \int_{\beta_{i}}^{1-\alpha}
d\beta \alpha\beta (s-\widetilde{m}^2_Q)(3s-\widetilde{m}^2_Q)\nonumber  \\
&& -\frac{1}{2048 \pi^4} \int_{\alpha_{i}}^{\alpha_{f}}d\alpha
\int_{\beta_{i}}^{1-\alpha}
d\beta (1-\alpha-\beta)^2 (s-\widetilde{m}^2_Q)(5s-3\widetilde{m}^2_Q)\nonumber  \\
&&+\frac{3m_Q^2}{512 \pi^4} \int_{\alpha_{i}}^{\alpha_{f}}d\alpha
\int_{\beta_{i}}^{1-\alpha}
d\beta(s-\widetilde{m}^2_Q)\nonumber  \\
&&- \frac{m_Q^2}{1024 \pi^4} \int_{\alpha_{i}}^{\alpha_{f}}d\alpha
\int_{\beta_{i}}^{1-\alpha}
d\beta\left[ \frac{\alpha}{\beta^2}+\frac{\beta}{\alpha^2}\right] (1-\alpha-\beta)^2(2s-\widetilde{m}^2_Q)\nonumber  \\
&&+ \frac{3m_Q^2}{1024 \pi^4} \int_{\alpha_{i}}^{\alpha_{f}}d\alpha
\int_{\beta_{i}}^{1-\alpha}
d\beta\left[ \frac{1}{\alpha^2}+\frac{1}{\beta^2}\right] (1-\alpha-\beta)^2(s-\widetilde{m}^2_Q)\nonumber  \\
&&- \frac{m_Q^4}{1024 \pi^4} \int_{\alpha_{i}}^{\alpha_{f}}d\alpha
\int_{\beta_{i}}^{1-\alpha} d\beta
\left[ \frac{1}{\alpha^3}+\frac{1}{\beta^3}\right] (1-\alpha-\beta)^2 \nonumber  \\
&&- \frac{m_s\langle\bar{s} s\rangle}{128 \pi^2}
\int_{\alpha_{i}}^{\alpha_{f}}d\alpha \int_{\beta_{i}}^{1-\alpha}
d\beta\left[ 3+s\delta(s-\widetilde{m}^2_Q)\right] \nonumber  \\
&&- \frac{m_sm_Q^2\langle\bar{s} s\rangle}{128 \pi^2M^2}
\int_{\alpha_{i}}^{\alpha_{f}}d\alpha \int_{\beta_{i}}^{1-\alpha}
d\beta\left[ \frac{\alpha}{\beta^2}+\frac{\beta}{\alpha^2}\right] s\delta(s-\widetilde{m}^2_Q)\nonumber  \\
&&- \frac{m_sm_Q^4\langle\bar{s} s\rangle}{128 \pi^2M^2}
\int_{\alpha_{i}}^{\alpha_{f}}d\alpha \int_{\beta_{i}}^{1-\alpha}
d\beta\left[ \frac{1}{\alpha^3}+\frac{1}{\beta^3}\right] \delta(s-\widetilde{m}^2_Q)\nonumber  \\
&&+ \frac{3m_sm_Q^2\langle\bar{s} s\rangle}{128 \pi^2 }
\int_{\alpha_{i}}^{\alpha_{f}}d\alpha \int_{\beta_{i}}^{1-\alpha}
d\beta\left[ \frac{1}{\alpha^2}+\frac{1}{\beta^2}\right]
\delta(s-\widetilde{m}^2_Q)\, ,
\end{eqnarray}

\begin{eqnarray}
\rho^B_{\langle GG\rangle}(s)&=& \frac{m_s\langle\bar{s}g_s\sigma
Gs\rangle}{576 \pi^2} \int_{\alpha_{i}}^{\alpha_{f}}d\alpha
\left[2+\frac{s}{M^2}\right]\delta(s-\widetilde{\widetilde{m}}^2_Q)\nonumber  \\
&& +\frac{\langle\bar{s}s\rangle^2}{288}
\int_{\alpha_{i}}^{\alpha_{f}}d\alpha
\left[2+\frac{s}{M^2}\right]\delta(s-\widetilde{\widetilde{m}}^2_Q)\nonumber\\
&&- \frac{m_sm_Q^2\langle\bar{s}g_s\sigma Gs\rangle}{576 \pi^2M^2}
\int_{\alpha_{i}}^{\alpha_{f}}d\alpha  \left[
\frac{1-\alpha}{\alpha^2}+\frac{\alpha}{(1-\alpha)^2}\right]
\left[1-\frac{s}{M^2}\right]\delta(s-\widetilde{\widetilde{m}}^2_Q)\nonumber  \\
&&+ \frac{m_sm_Q^4\langle\bar{s}g_s\sigma Gs\rangle}{576 \pi^2M^4}
\int_{\alpha_{i}}^{\alpha_{f}}d\alpha  \left[
\frac{1}{\alpha^3}+\frac{1}{(1-\alpha)^3}\right]
\delta(s-\widetilde{\widetilde{m}}^2_Q)\nonumber  \\
&&- \frac{m_Q^2\langle\bar{s}s\rangle^2}{288M^2}
\int_{\alpha_{i}}^{\alpha_{f}}d\alpha  \left[
\frac{1-\alpha}{\alpha^2}+\frac{\alpha}{(1-\alpha)^2}\right]
\left[1-\frac{s}{M^2}\right]\delta(s-\widetilde{\widetilde{m}}^2_Q)\nonumber  \\
&&+ \frac{m_Q^4\langle\bar{s}s\rangle^2}{288M^4}
\int_{\alpha_{i}}^{\alpha_{f}}d\alpha  \left[
\frac{1}{\alpha^3}+\frac{1}{(1-\alpha)^3}\right]
\delta(s-\widetilde{\widetilde{m}}^2_Q)\nonumber  \\
&&- \frac{m_sm_Q^2\langle\bar{s}g_s\sigma Gs\rangle}{192 \pi^2M^2}
\int_{\alpha_{i}}^{\alpha_{f}}d\alpha  \left[
\frac{1}{\alpha^2}+\frac{1}{(1-\alpha)^2}\right]
\delta(s-\widetilde{\widetilde{m}}^2_Q)\nonumber  \\
&&- \frac{m_Q^2\langle\bar{s}s\rangle^2}{96M^2}
\int_{\alpha_{i}}^{\alpha_{f}}d\alpha  \left[
\frac{1}{\alpha^2}+\frac{1}{(1-\alpha)^2}\right]
\delta(s-\widetilde{\widetilde{m}}^2_Q) \,.
\end{eqnarray}

\section*{Acknowledgements}
This  work is supported by National Natural Science Foundation of
China, Grant Number 10775051, and Program for New Century Excellent
Talents in University, Grant Number NCET-07-0282.

\end{document}